\documentclass[aps, prl, twocolumn, showpacs, superscriptaddress]{revtex4-1}
\usepackage{color}
\usepackage{graphicx}
\usepackage{amsmath, amssymb, amsbsy}

\DeclareGraphicsExtensions{{.eps},{.png},{.pdf}}

\def\bra#1{\mathinner{\langle{#1}|}}
\def\ket#1{\mathinner{|{#1}\rangle}}
\def\abs#1{\ifmmode
                \left \vert #1 \right \vert%
           \else
                $\left \vert #1 \right \vert$%
           \fi}
\DeclareMathOperator{\Tr}{Tr}

\begin{document}

\title{Improving quantum annealing by engineering the coupling to the environment}
\author{Mojdeh S. Najafabadi} 
\affiliation{Dodd-Walls Centre for Photonic and Quantum Technologies,
             Department of Physics, University of Otago, Dunedin, New Zealand}
\affiliation{Max-Planck Institute for the Science of Light, Erlangen, Germany}             

\author{Daniel Schumayer}
\affiliation{Dodd-Walls Centre for Photonic and Quantum Technologies,
             Department of Physics, University of Otago, Dunedin, New Zealand}

\author{Chee Kong Lee}
\affiliation{Tencent America, Palo Alto, California 94301, USA}

\author{Dieter Jaksch}
\affiliation{Clarendon Laboratory, Department of Physics,
             University of Oxford, OX1 3PU, United Kingdom}
             
\author{David A. W. Hutchinson}  \email{david.hutchinson@otago.ac.nz}
\affiliation{Dodd-Walls Centre for Photonic and Quantum Technologies, 
             Department of Physics, University of Otago, Dunedin, New Zealand}
\affiliation{Clarendon Laboratory, Department of Physics,
             University of Oxford, OX1 3PU, United Kingdom}
           
\begin{abstract}
 {A large class of optimisation problems can be mapped to the Ising model where all details are encoded in the coupling of spins. The task of the original mathematical optimisation is then equivalent to finding the ground state of the corresponding spin system which can be achieved via quantum annealing relying on the adiabatic theorem. Some of the inherent disadvantages of this procedure can be alleviated or resolved using a stochastic approach, and by coupling to the external environment. We show that careful engineering of the system-bath coupling at an {\emph{individual}} spin level can further improve annealing.}
\end{abstract}
\date{\today} \maketitle

\section{Introduction}

Quantum computation is expected to outstrip its classical counterpart in certain mathematical algorithms such as integer factorization \cite{Shor1994}, in data searches \cite{Grover1997}, and also in large scale optimisation \cite{Lloyd1996}, especially if the target function corresponds directly to the quantum-mechanical description of the system, e.g., ground state \cite{Abrams1999} or correlation structure \cite{Abrams1997}. Hence, should a challenging combinatorial problem be mappable onto a quantum system, a quantum simulator may indirectly solve the original problem in reasonable time.

A prototypical and suitable physical system is an Ising spin system, which suits experimental realisation, and all details of the hypothetical problem (e.g., the traveling salesman problem \cite{Cook2011}) are encoded in the interaction strengths between spins, $J_{ij}$. By finding the ground state, for given $J_{ij}$, one solves the original problem \cite{Farhi2000, Martonak2004, Pirchi2011, Lucas2014, Srinivasan2018, Warren2019, Dan2020, Tan2021}. Of course, finding the ground state is itself an NP-hard problem for a large Ising system \cite{Barahona1982}, in general. However, a potentially feasible strategy is to let Nature find the solution in a quick experimental protocol.

One approach --exploiting Kato's theorem \cite{Kato1950}-- is to prepare the system in the ground state of a simple Hamiltonian, and tune this Hamiltonian to the one whose ground state we are actually seeking. Kato's adiabatic theorem states that if a self-adjoint operator has an isolated eigenvalue with a potentially degenerate eigenspace and this eigenvalue does not split under a smooth self-adjoint perturbation, then there is a unitary transformation which transforms the unperturbed eigenspace to the corresponding perturbed eigenspace. In other words this theorem guarantees that a ground state of an unperturbed Hamiltonian can be evolved into the ground state of another Hamiltonian as long as the change of Hamiltonians happens smoothly enough.

One may apply a strong magnetic field in the $x$-direction to align all spins initially, then turning off the field slowly, and eventually measuring the spins in the $z$-directions. Unfortunately, in this process the energy levels of the instantaneous Hamiltonian usually become very close, and hence the driving must slow down to remain adiabatic (viz. critical slowing down \cite{Suzuki1970, Salamon1978, Binder2012, Tredicce2004, Biroli2010}). Alternatively, proceeding in a finite time elevates the probability of the system being excited. As real quantum systems always interact with their environment, one may make a virtue of necessity, and allow the spins to couple to the thermal or quantum fluctuations of this external bath \cite{Magalinskii1959, Zwanzig1973, Caldeira1981, Buca2019}. Repeating the experiment a number of times yields an ensemble of final states from which we can pick the lowest energy configuration since calculating the energy for a given configuration is simple. In this approach, the likelihood of `not finding the true ground state' diminishes exponentially with the number of trials \cite{Johnson2011}.

Engineering the interaction with the environment can improve quantum annealing \cite{Amin2008, Ashhab2014, Kechedzhi2016, Keck2017, Arceci2017, Arceci2018, Passarelli2018, Lee2019}, and such active design, perhaps individually to each qubit, forms the heart of this paper. Inspired by energy transport optimization through site-dependent coupling to the environment in photosynthetic systems \cite{Oh2019, Oh2020}, we propose the use of site-specific dissipation. We investigate this approach by evolving an Ising system via the Bloch-Redfield master equation. Within its approximations quantum annealing can achieve high efficiency \cite{Gaspard1999, Nalbach2009, May2011, Nalbach2014, Yoshimura2015, Javanbakht2015, Watabe2020}. We show that online learning and adjustment of the individual coupling of each qubit to the environment indeed increases the probability of success, and that the annealing time is also an important factor in the protocol.

\section{Model}

\begin{figure}[t!]
    \includegraphics[width=80mm]%
                    {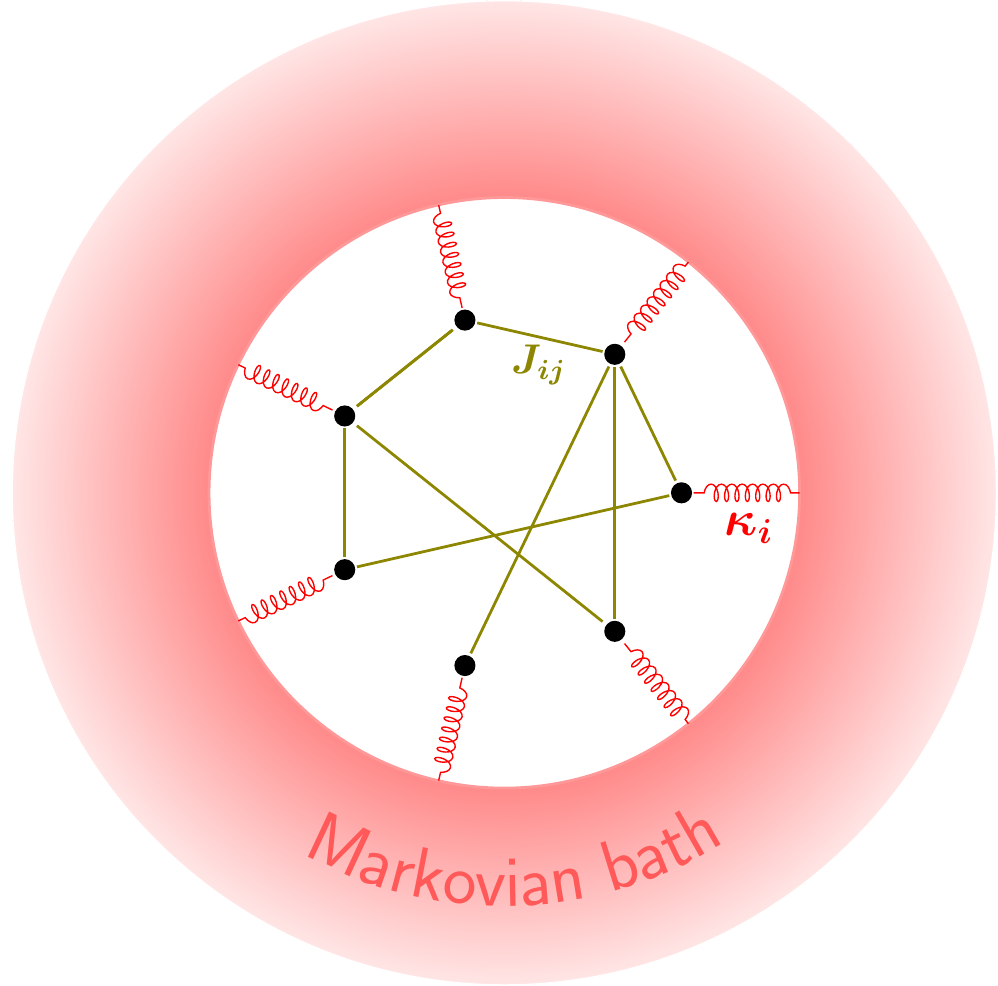}
    \caption{\label{fig:Schematics}
             Schematics of 7 qubits (black dots) interacting with each other with strengths $J_{ij}$ (olive solid lines) and with a Markovian bath (red springs) with couplings $\kappa_{i}$.
            }
\end{figure}

Let us consider a collection of $N$ identical, interacting qubits arranged at the vertices of an undirected graph, $G$, as depicted in Fig.~\ref{fig:Schematics}:
\begin{equation*}
    H_{G}
    = 
    - m_{0}^{\phantom{z}}\sigma_{0}^{z}
    -\frac{1}{2}
    \sum_{\substack{i, j=1}}^{N}
        {J_{ij}^{\phantom{z}} \sigma_{i}^{z} \sigma_{j}^{z}}
\end{equation*}
with ferromagnetic couplings ($J_{ij} < 0$). A small field, $m_{0} = \mu B_{0}$, is pinning a single qubit, $\sigma_{0}$, in order to resolve the degeneracy of the ground state. The system is controlled by time-dependent external magnetic field
\begin{equation*}
    H_{\text{ext}}
    =
    - \mu m_{x,0} e^{-t/\tau} \sum_{i}^{N} \sigma_{i}^{x},
\end{equation*}
where $\tau$ is the annealing time, and $m_{x,0}$ is chosen to overwhelm all other terms at the start of the experiment.

We consider two cases: (a) uniform coupling ($J_{ij} = J$), and (b) random couplings, where every $J_{ij}$ are drawn from a normal distribution $\mathcal{N}(-2, 0.1)$. We distinguish three classes of graphs: random non-complete graphs, a linear graph and a complete graph. The latter two are unique by their adjacency structure and form the extremes of connected graphs: a linear graph has the least, while the fully connected graph the most edges while being connected. In experiments we envisage a system somewhere in between these cases. In order to investigate these `more typical' non-complete graphs, we sampled connected graphs for which half of all possible edges were missing. An exhaustive simulation could explore random connected graphs with $k$ edges, but uniformly sampling this set of graphs is not a trivial task \cite{Bender1978}.

The Ising system is brought into contact with an infinitely large bath in thermal equilibrium modelled as
\begin{equation*} 
    H_{\text{bath}}
    = 
     \sum_{\lambda}
         {\hbar \omega_{\lambda}^{\phantom{\dag}}
          b_{\lambda}^{\dag} b_{\lambda}^{\phantom{\dag}}
         }.
\end{equation*} 
The summation runs over the oscillator modes, while $b_{\lambda}^{\dag}$ and $b_{\lambda}^{\phantom{\dag}}$ are the bosonic creation and annihilation operators for mode ${\lambda}$. Finally, the qubit-environment interaction is
\begin{equation*} \label{eq:InteractionHamiltonian}
    H_{\text{int}} = \sum_{i} A_{i} \otimes B,
\end{equation*}
where we opted for $A_{i} = \kappa_{i}(\sigma^{+}_{i} + \sigma^{-}_{i})$ with site-specific coupling $\kappa_{i}$, and $B = \sum_{\lambda} (b_{\lambda}^{\dag} + b_{\lambda}^{\phantom{\dag}})$ as operators of the quantum system and of the environment, respectively. For the sake of simplicity, although unphysical, we assume $\kappa_{i}$ being independent of energy, i.e., a qubit is coupled to all bath modes equally. 

The Hamiltonian, $H = H_{G} + H_{\text{ext}} + H_{\text{bath}} + H_{\text{int}}$, governs the time evolution $\dot{\rho}_{\text{total}}
= -i \lbrack H_{\text{int}}, \rho_{\text{total}} \rbrack$ in the interaction picture. As we are interested in the dynamics of the qubits, we trace over the bath states and obtain
\begin{equation}\label{eq:TimeEvolutionOfDensityOperator}
    \dot{\rho}(t)
    =
    - 
    \int_{0}^{t} ds
    \Tr_{\text{bath}}
    \left \lbrack
        H_{\text{int}}(t),
        \bigl \lbrack
            H_{\text{int}}(s), \rho_{\text{total}}(s)
        \bigr \rbrack
    \right \rbrack.
\end{equation}
In solving Eq.~\eqref{eq:TimeEvolutionOfDensityOperator} we make the standard assumptions and follow Refs.~\cite{Tannoudji1992, Breuer2002, Johansson2013}. The qubit-bath coupling is weak (Born approximation). The bath is initially uncorrelated. The density matrix can be factorized (i.e., no appreciable correlation between system and bath) which is kept throughout the evolution, hence $\rho_{\text{total}}(t) \approx \rho(t) \otimes \rho_{\text{bath}}$, where $\rho_{\text{bath}} \propto \exp{\!\left (-\beta H_{\text{bath}} \right )}$ is the canonical density matrix at inverse temperature $\beta$. Finally, the bath is assumed to be Markovian, which allows us to change the integration limit in Eq.~\eqref{eq:TimeEvolutionOfDensityOperator} from $t$ to $\infty$ leading to
\begin{equation} \label{eq:MarkovianMasterEquation}
    \dot{\rho}(t)
    =
    \sum_{j k} \int_{0}^{\infty} ds
        \Bigl \lbrack
            A_{j}(t); A_{k}(s) \rho(s) 
        \Bigr \rbrack
        C_{\text{bath}}
        + \text{h.c.}
\end{equation}
Here $C_{\text{bath}} = \Tr_{\text{bath}}(B(t) B(t') \rho_{\text{bath}})$ is the bath time-correlation function. Using the energy eigenbasis of $H_{\text{sys}} = H_{G} + H_{\text{ext}}$, e.g., $\bra{a} A(t) \ket{b} = A_{ab} e^{i\omega_{ab} t}$, we transform Eq.~\eqref{eq:MarkovianMasterEquation} into a matrix equation
\begin{equation} \label{eq:BlochRedfieldMasterEquation_MatrixForm}
    \dot{\rho}_{ab}
    \cong
    -i \omega_{ab} \rho_{ab}
    - \sum_{cd}{R_{abcd} \rho_{cd}}.
\end{equation}
We discard fast oscillatory terms and keep only those terms in the summation for which $\abs{\omega_{ab} - \omega_{cd}}^{-1} \ll \tau_{\text{relax}}$. For clarity, we introduced the Bloch-Redfield tensor $R_{abcd} = -\frac{1}{2} \sum_{jk}{\!\bigl ( \delta_{bd}^{\phantom{\beta}} r_{ac}^{jk} + \delta_{ac}^{\phantom{\beta}} r_{db}^{jk} \bigr )}$ with
\begin{align*}
    r_{ac}^{jk}
    &=
    \sum_{n}{A_{an}^{j} A_{nc}^{k} S(\omega_{cn}) -
             A_{ac}^{j} A_{db}^{k} S(\omega_{ca})}
    \\
    r_{db}^{jk}
    &=
    \sum_{n}{A_{dn}^{j} A_{nb}^{k} S(\omega_{dn}) -
             A_{ac}^{j} A_{db}^{k} S(\omega_{db})}.
\end{align*}
Here $S (\omega)= \int_{-\infty}^{\infty} d\tau e^{i\omega t} C(t) \cong 2\pi J(\omega) (1 + \overline{n} (\omega))$ is the noise power-spectrum of the bath. The function $J(\omega)=\eta \omega e^{-\omega/\omega_{c}}$ is the Ohmic spectral density function with a cut-off frequency $\omega_{c}$ and a dimensionless parameter $\eta$~\cite{Lidar2020}, while $\overline{n}(\omega) =(e^{\beta \hbar \omega}-1)^{-1}$ is the mean oscillator number in mode $\omega$ at inverse temperature $\beta$. Usually, for cold enough baths the relaxation rate is considered to be Ohmic \cite{Yan2016}, and, alternatively, an ensemble of coherent two-level systems can also reproduce Ohmic excitation spectrum \cite*{Astafiev2004, Shnirman2005}.

We emphasise that the bath correlation time, $\tau_{\text{bath}}$, must be short enough, such that
\begin{equation} \label{eq:BornMarkovCondition}
    \tau_{\text{bath}} \ll \tau_{\text{relax}}.
\end{equation}
In other words the fast oscillating quantities average out, and the terms with frequency $\omega_{ab} - \omega_{cd}$ will not give any significant contribution to the system evolution by $t'$ such that $\abs{\omega_{ab} - \omega_{cd}}^{-1} \ll t' \ll \tau_{\text{relax}}$. This condition also sets a limit on the time-scale of the simulation \cite{Johansson2012, Johansson2013}, as beyond $\tau_{\text{relax}}$ the observables do not change any more.

As a crude approximation $\tau_{\text{relax}} \propto \kappa^{-2}$ \cite{Albash2012, Cattaneo2019} while
\begin{equation*}
    \tau_{\text{bath}}
    \cong
    \max{ \left ( \frac{2\pi}{\omega_{c}},  \beta \right )}.
\end{equation*}
Hence as long as $\tau_{\text{bath}} \ll \tau_{\text{relax}}$ is fulfilled, Eq.~\eqref{eq:BlochRedfieldMasterEquation_MatrixForm} should be an adequate description. The suitable $\beta$ also depends on the graph and we use $\beta=1.3$ for the linear graph, 2 for the random non-complete graph, and 2.2 for the complete graph. These $\beta$ values lead to $\kappa_{i} \lesssim 2$.

\section{Results}

Fidelity is a central measure in quantum computation and an upper bound can be calculated as \cite{Jozsa1994, Miszczak2008} 
\begin{equation} \label{eq:MiszczakFidelity}
    F(\rho_{1}, \rho_{2})
    =
    \Tr{(\rho_{1} \rho_{2})} +
    \sqrt{1 - \Tr{(\rho_{1}^{2})}}\,
    \sqrt{1-\Tr{(\rho_{2}^2)}},
\end{equation}
where ${\rho}_{1}$ and $\rho_{2}$ are any two density operators. In the following $\rho_{1}$ is fixed by the instantaneous ground state, while $\rho_{2} = \rho(t)$ is the instantaneous density operator. We also follow the time evolution of the instantaneous energy of the system, $\varepsilon(t) = \Tr \left ( \rho H_{\text{sys}}(t) \right )$, compared to $\Tr \left ( \rho H_{\text{sys}}(t=\infty) \right )$. 
\begin{figure}[t!]
    \includegraphics[width=82mm]%
                    {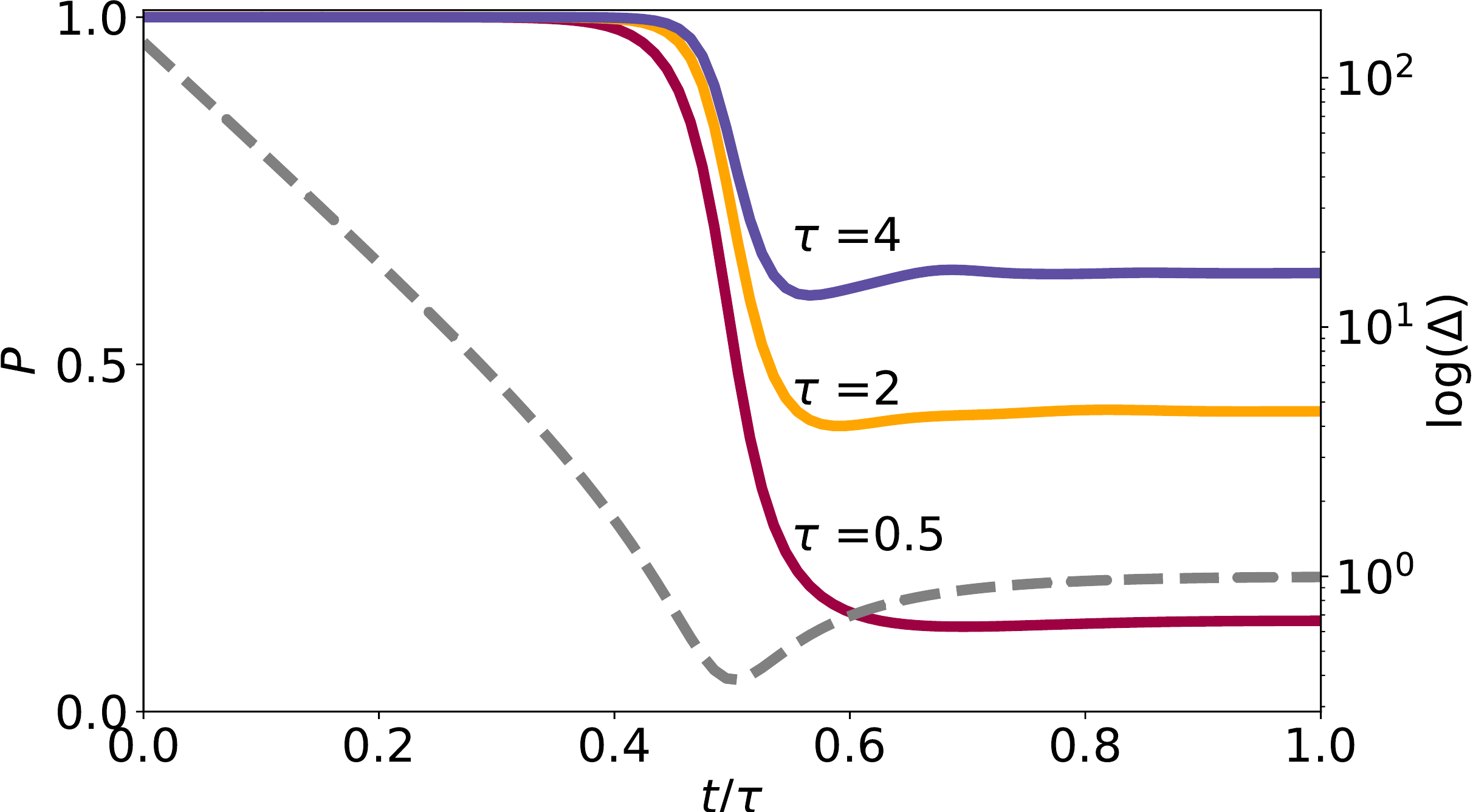}
    \caption{\label{fig:PopulationFidelityForDifferentAnnealingTimes}
             The time dependence of $P$ is plotted for an isolated 7-qubit chain for different $\tau$ (solid lines). The dashed line is the logarithm of the energy gap, $\Delta$, as a function of scaled time. Other parameters: $m_{0,x} = 140$, $J_{ij}=1$. 
            }
\end{figure}

As a benchmark, we prepare the 7-qubit chain in its ground state with constant $J=1$, and anneal with different $\tau$ without coupling to the environment.
The probability of being in the instantaneous ground state, $P$, together with the logarithm of the gap between the ground state and the first excited state (dashed line), $\log(\Delta)$, are shown in the Fig.~\ref{fig:PopulationFidelityForDifferentAnnealingTimes}. The gap initially diminishes rapidly, reaches its minimal value around $t/\tau \approx 0.5$, and then levels off at a value on the order of unity. Not surprisingly, around the $\min{\!(\Delta)}$ fidelity drops as different states start mixing. However, for slower annealing ($\tau=2$, 4) scattering into the excited states is less, and higher $P$ values are maintained. As expected $\tau$ plays an important role \cite{Albash2012}. Although short annealing time reduces probability of being in the ground state, we do aim to anneal fast, and hope that coupling to the environment helps the system to release its surplus energy and regain some of its lost population in the ground state. 
\begin{figure}[b!]
    \includegraphics[width=82mm]{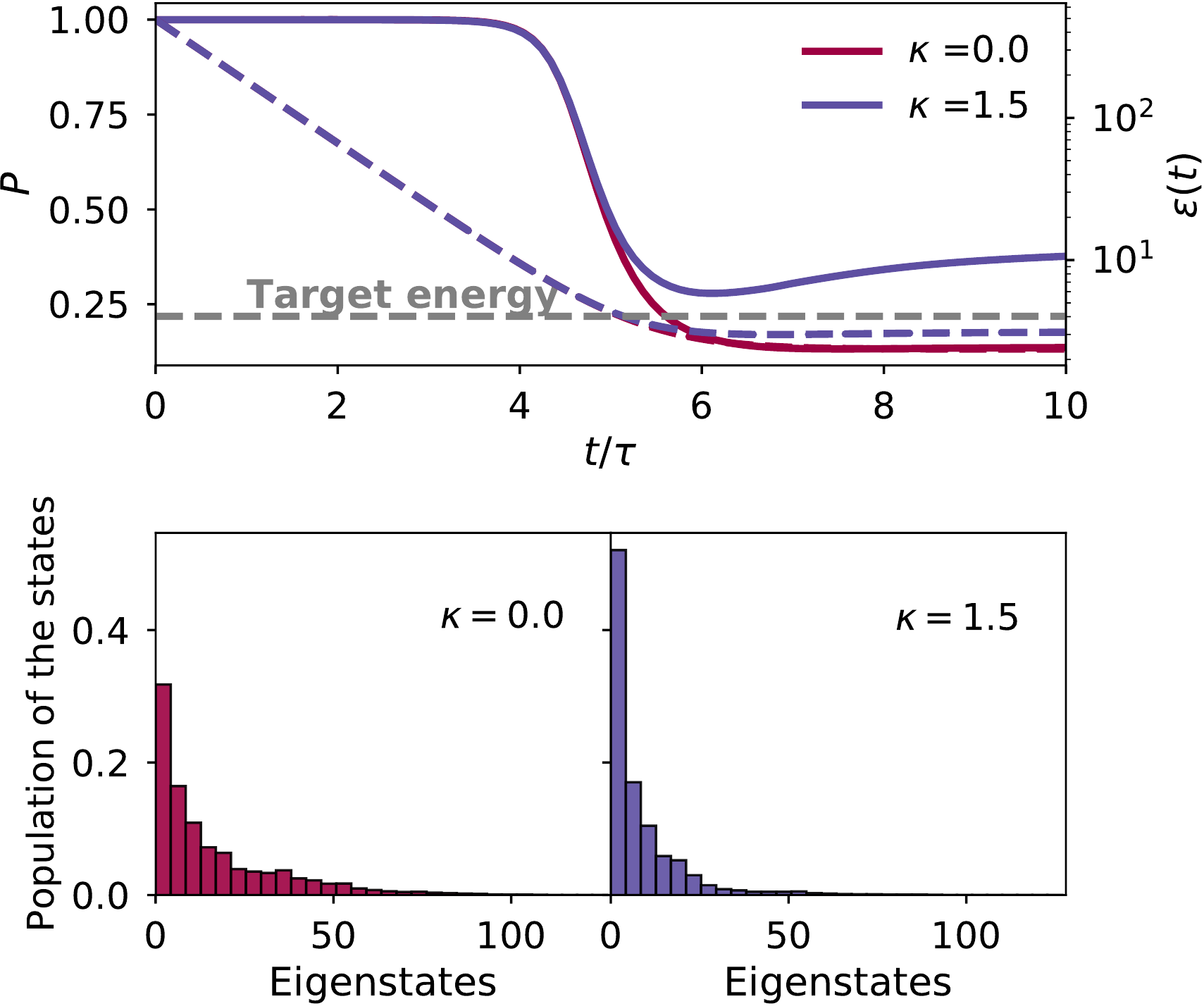}
    \caption{\label{fig:Probability_OpenClosed}
             (Top) Probability of being in the ground state (solid lines) over time for isolated ($\kappa=0$) and open system ($\kappa=1.5$) for 7 qubits. The instantaneous energy (dashed lines) of both systems approaching the target energy. (Bottom) Populations of all $2^{7}$ energy eigenstates at the end of the experiment for both systems.
            }
\end{figure}

Let us compare the population of the ground state for open and closed system with fixed $\tau = \tfrac{1}{2}$. Fig~\ref{fig:Probability_OpenClosed} shows $\varepsilon(t)$ and $P$ for closed ($\kappa=0$) and open ($\kappa=1.5$) systems. The open system coupled weakly to the environment ends up in the instantaneous ground state with higher probability, and simultaneously the instantaneous energy, $\varepsilon(t)$ approaches the target energy. 
The bottom panel of Fig.~\ref{fig:Probability_OpenClosed}, compares population of states of each system: the ground state population remains much higher for a weakly coupled system than for the isolated system. Concluding, weak coupling to the environment may help the system to evolve into its ground state with higher probability.

Next we vary $\kappa_{i}$ to maximize the population of the ground state at the end of the simulation, $P_{\text{fin}}$.~\footnote{Each simulation consists of two parts: one solves the Bloch-Redfield master equation and measure the fidelity between two given states, and the other optimizes the target function $y = 1 - \max{\!\bigl ( \text{fidelity} \bigr)}^{2}$ with the constraint $0 < \kappa$ . We rely on the {\texttt{python3}} optimization module {\texttt{scipy.optimize}} implementing the Nelder-Mead algorithm \cite{Nelder1965}. A run starts with an initial guess for $\kappa$, measures $F$ as in Eq.~\eqref{eq:MiszczakFidelity}, and then $y$ is evaluated.}. First, we consider a uniform coupling, $\kappa_{i} = \kappa^{\text{opt}}$, while later we allow for site-dependence, $\kappa_{i} = \kappa_{i}^{\text{opt}}$.
\begin{figure}[b!]
    \includegraphics[width=82mm]%
                    {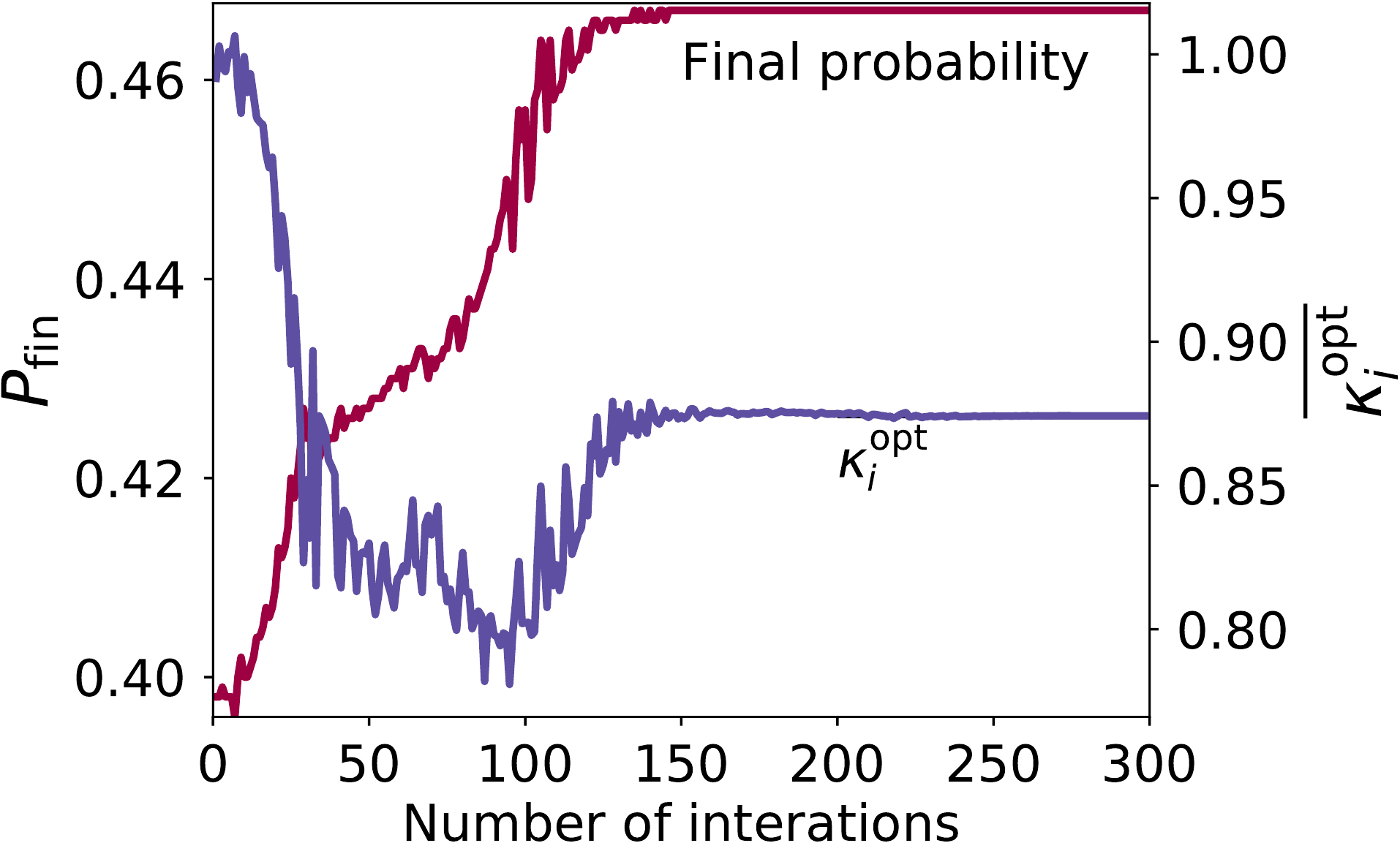}
    \caption{\label{fig:FinalProbability_AverageCoupling}
             The increasing purple line shows $P_{\text{fin}}$, while the blue line corresponds to the average coupling strength, $\overline{\kappa_{i}^{\text{opt}}}$, in each step of the optimization process.
            }
\end{figure}
The evolution of $P_{\text{fin}}$ and the average coupling strengths, $\overline{\kappa}$, are depicted in Fig.~\ref{fig:FinalProbability_AverageCoupling}. We pick $\lbrace \kappa_{i} \rbrace$ corresponding to the highest $P_{\text{fin}}$.

\begin{figure}[b!]
    \includegraphics[width=82mm]%
                    {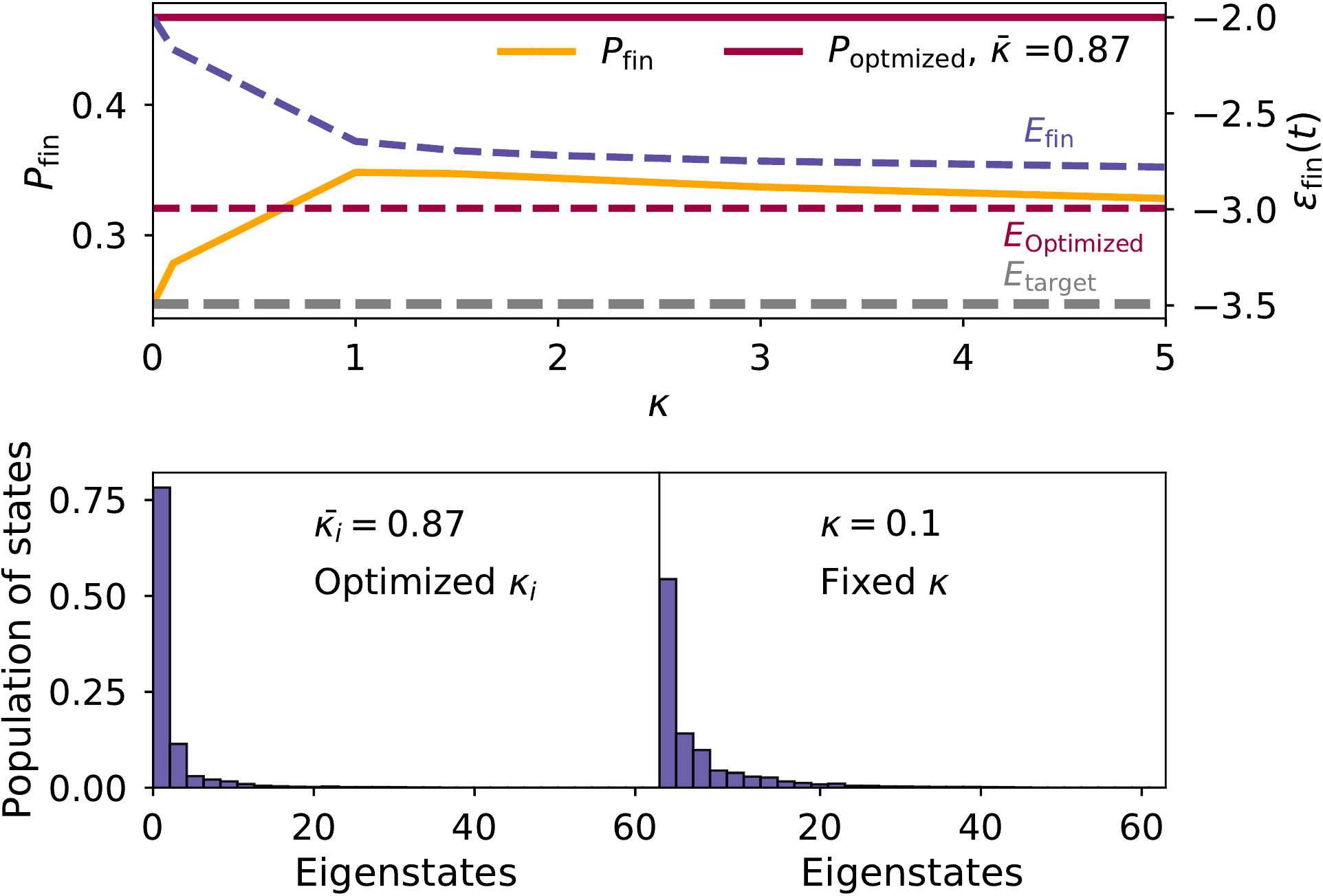}
    \caption{\label{fig:Probability_FixedAndOptimized}
             The probability of being in the ground state at the end of the simulation for a non-complete graph for random $J_{ij}$. The coupling to the bath is either kept constant or it is optimised. The histograms show the populations of each energy eigenstates at the end of the evolution. Other parameters: $N=6$, $\beta = 2.2$, and $\omega_{c} = 30$.
            }
\end{figure}

Fig~\ref{fig:Probability_FixedAndOptimized} shows $P_{\text{fin}}$ for fixed non-optimal coupling and for $\kappa_{i}^{\text{opt}}$. It is apparent that $P_{\text{fin}} < P_{\text{optimized}}$ for a non-optimal $\kappa$, and also the final energy of the optimised case is closer to the target energy (the true ground state energy) than for the non-optimised system. 

One may ponder upon the stability of the ground state population above some $\kappa$. We have simulated the time evolution for both non-optimized and optimized system-bath couplings. 
\begin{figure}[t!]
    \includegraphics[width=82mm]%
                    {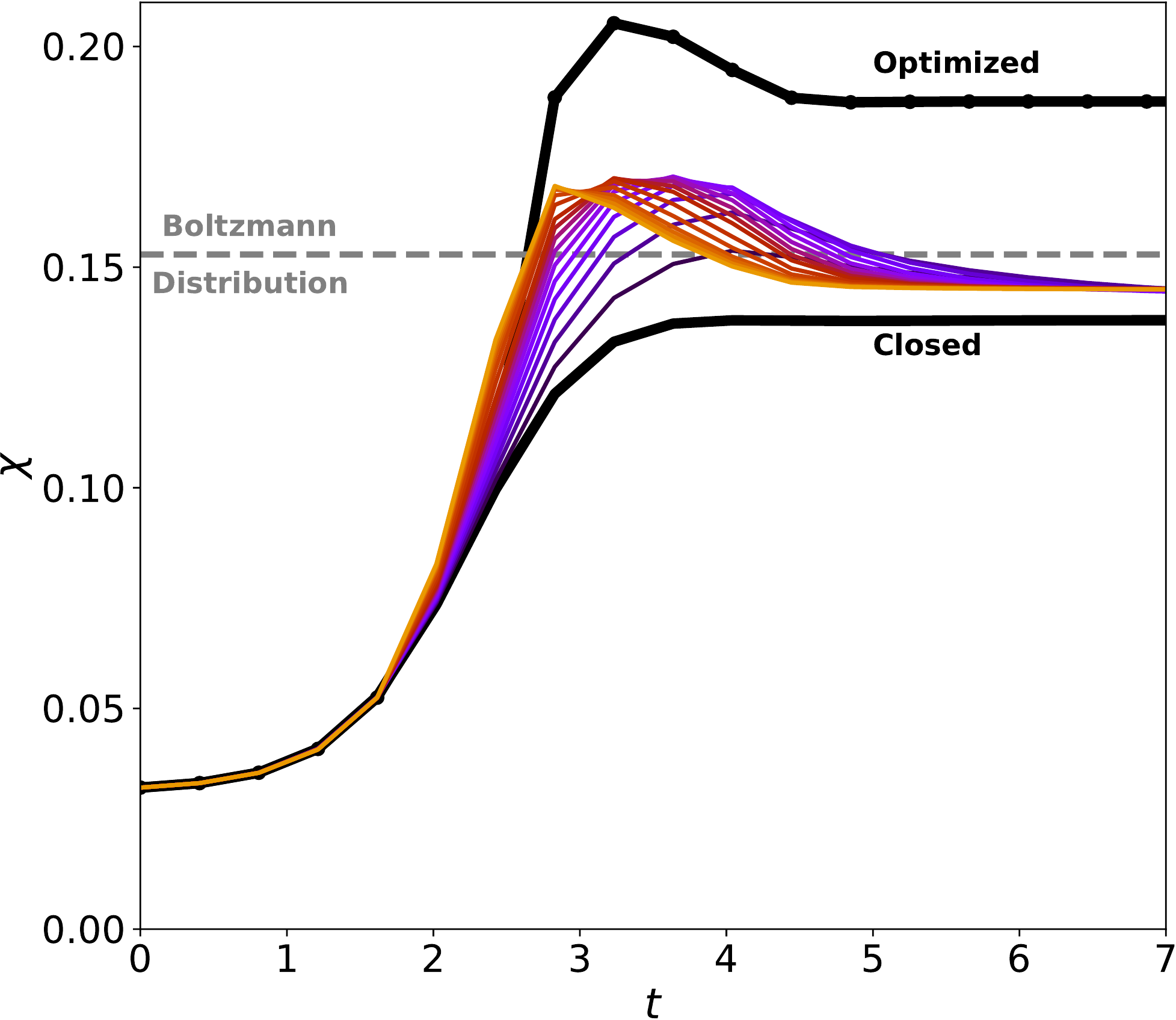}
    \caption{\label{fig:Determine_tmax}
             Time evolution of $\chi$ for a non-complete random graph for 5 qubits and fixed $\tau=0.5$. The gray dashed line shows the probability expected for the final ground state in thermal equilibrium. The coloured solid lines are for non-optimized open systems with $0.1 \le \kappa_{i} \le 2$. The top solid black line is for the optimised system while the bottom solid black line corresponds to an isolated system.
            }
\end{figure}
Fig~\ref{fig:Determine_tmax} shows the results for a non-complete graph with 5 nodes coupled to a bath at inverse temperature $\beta=1.2$ with $0.1 \le \kappa_{i} \le 2$. If the annealing is stopped at $t \sim 3.2$, one attains the highest $\chi = F^{2}(\rho(t), \rho_{\text{gs}})$ which we interpret as the probability of being in the instantaneous ground state. If the process continues, the system ends up with a smaller population in the ground state than expected from Boltzmann's distribution. While the optimized system undergoes a qualitatively similar evolution, $\chi$ reaches a higher value even though there is a period (see $1.5 \lesssim t \lesssim 2.7$) when other couplings temporarily achieve higher $\chi$ values. 
\begin{figure}[b!]
    \includegraphics[width=82mm]%
                    {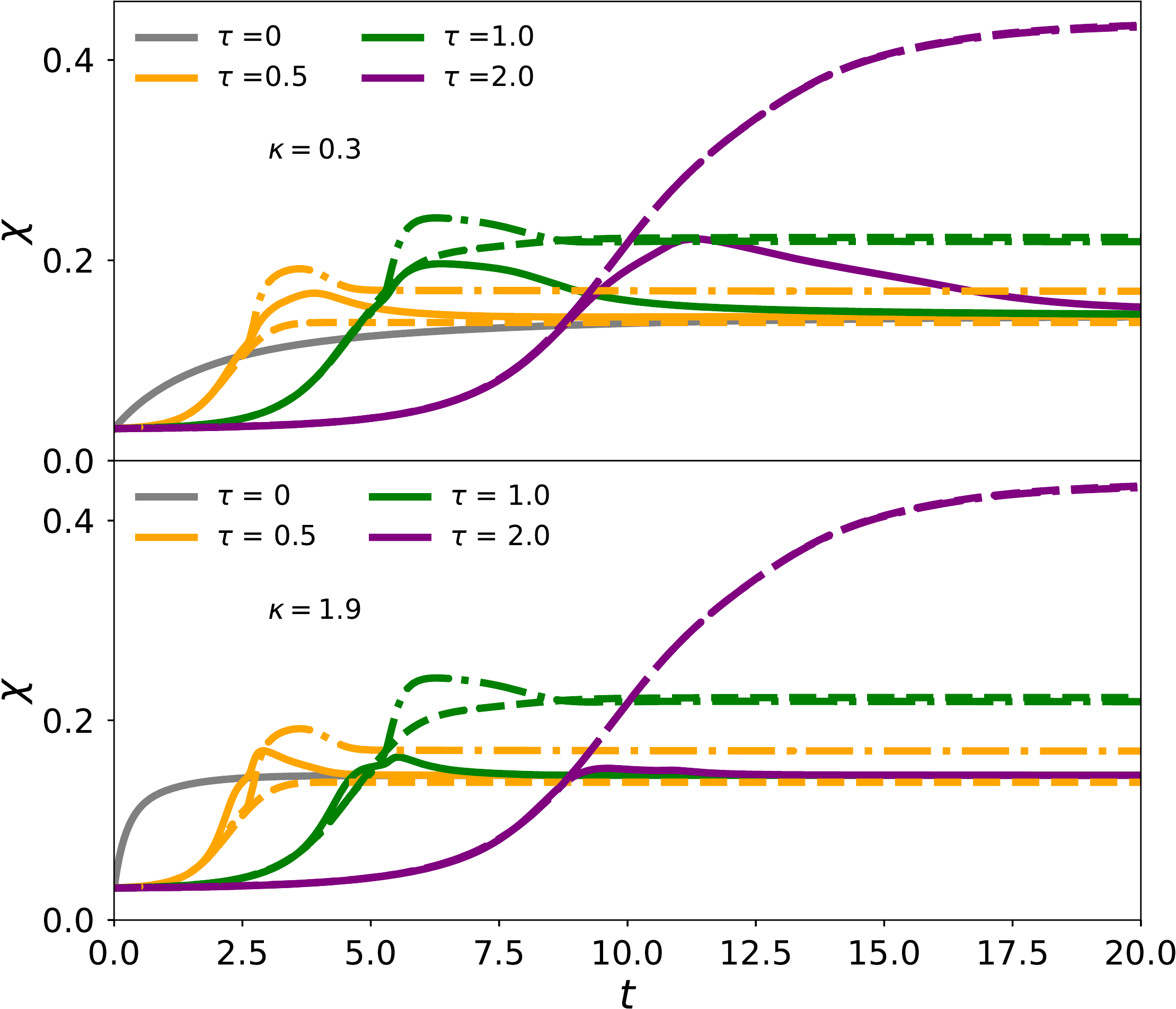}
    \caption{\label{fig:WithAndWithoutAnnealing_N5}
             Comparison of annealing and non-annealing processes by varying the annealing time, $\tau$. The solid lines are for open systems with $\kappa=0.3$ (top) and $1.9$ (bottom). The dash-dotted lines correspond to optimized $\kappa$, while the dashed lines are for an isolated system. 
            }
\end{figure}

We have also analysed whether annealing is better compared to suddenly turning $m(t)$ off at the very beginning of the time evolution after preparing the initial state. We chose a non-complete graph with 5 nodes, fixed $J_{ij}$ values, and varied $\tau$ from $0$ to $2$. The experiments are repeated for two coupling strengths $\kappa=0.3$ and 1.9. The results are plotted in Fig.~\ref{fig:WithAndWithoutAnnealing_N5} by line-triplets using the same colour for corresponding systems. One may draw three conclusions. First, all open annealed systems achieves higher $\chi$ values than the `no-annealing' protocol, although for strong environmental coupling the excess probability is small. Secondly, a slow protocol ($\tau=2$) may outperform asymptotically all quicker quenches and optimization does not improve $\chi$ appreciably. Finally, for all $\tau>0$ the time evolution of $\chi$ starts flat ($\tfrac{d}{dt} \chi \approx 0$), contrary to the `no-annealing' protocol which starts increasing immediately. Let us now focus on the coupling of medium strength, $\kappa=0.3$ (top panel of Fig.~\ref{fig:WithAndWithoutAnnealing_N5}). Comparing lines with identical colours, one may observe that their order changes as $\tau$ varies. For a quick protocol, $\tau = \tfrac{1}{2}$, the closed system performs the worst, an open system can improve on $\chi$ around $t \approx 3$, while optimizing $\kappa$ further increases $\chi$. For an intermediate value, $\tau = 1$, the closed system outperforms the open system for long enough experiments, but it is still worse than the optimized system. Driving the magnetic field even slower ($\tau=2$) the closed system and the optimised system performs identically, i.e., optimization does not improve the outcome. However, in reality no isolated system can be prepared, hence the conclusion remains: optimization improves $\chi$ and may worth the effort. 

The bottom panel of Fig.~\ref{fig:WithAndWithoutAnnealing_N5} corroborates our remarks. It is worth making two further comments. First, `no-annealing' protocol seems to reach its asymptotic value very quickly for stronger couplings, hence repeating it twice or thrice, instead of any smooth quenches, may be a better approach. Second, the maxima of $P$ for open systems with annealing protocols are significantly reduced and they perform only slightly better than the `no-annealing' protocol.

\begin{figure}
    \includegraphics[width=82mm]%
                    {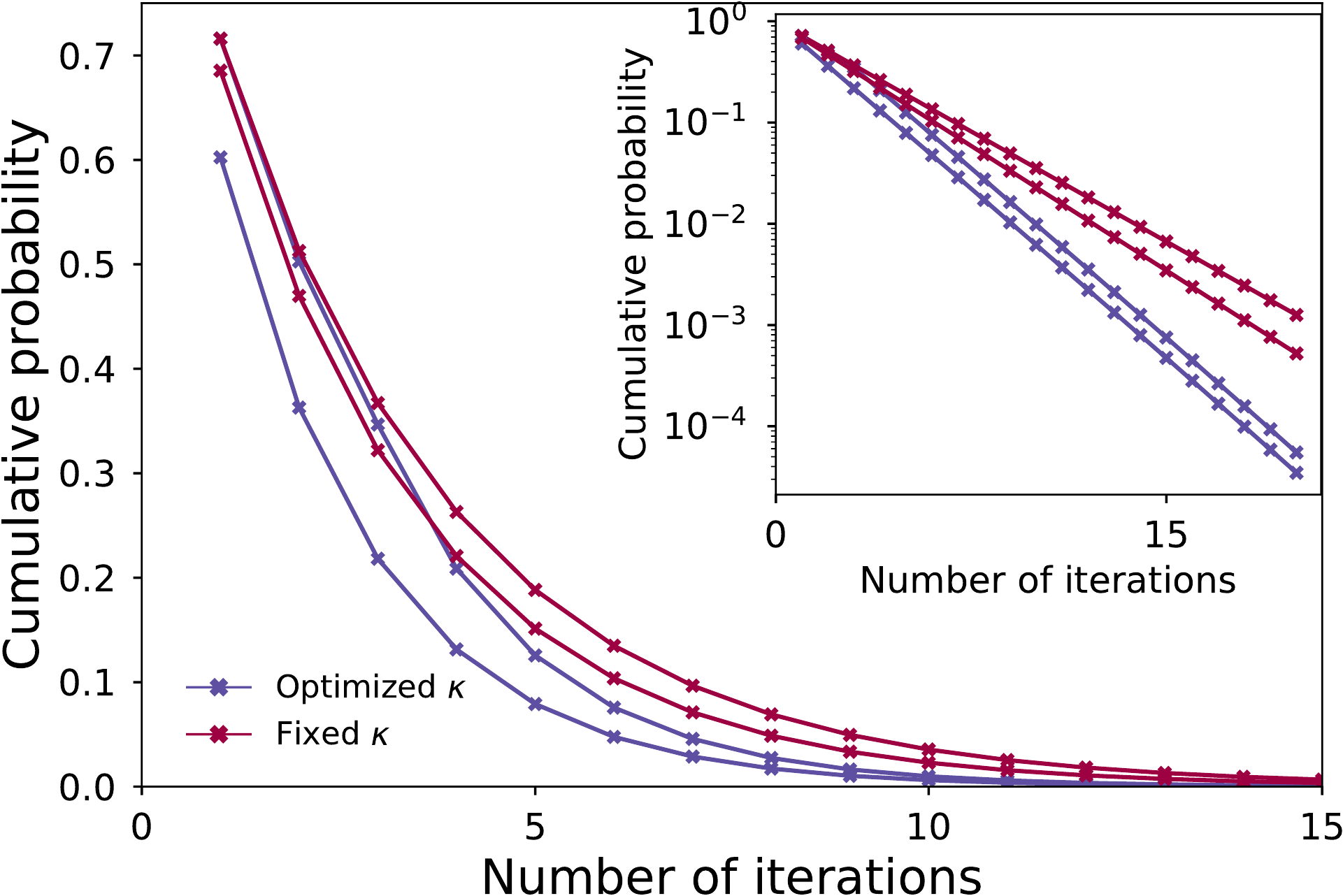} 
    \caption{\label{fig:CumulativeProbability}
             The cumulative probability, $P_{\text{c}}$, is shown for four runs, with random initial couplings. In two runs (red) these initial couplings are frozen and kept constant, while in the other runs $\kappa$ is optimized (blue) in each step. Even for the `bad' initial guesses the curves decrease faster. (Inset) Same data with logarithmic ordinate.
            }
\end{figure}

For both optimized and non-optimized cases we calculate the cumulative probability, $P_{\text{c}}$, of `not finding the true ground state in $n$ consecutive experiments' as
\begin{equation*}
    P_{\text{c}}(n) = \prod_{i=1}^{n}(1-P_{i}),
\end{equation*}
where $P_{i}$ is the probability of ending up in the ground state in the $i$\textsuperscript{th} experiment. In experiments, where the system-bath coupling is not optimized for, $P_{i}$ remains constant throughout, hence $P_{\text{c}}(n) = (1-P_{1})^{n}$. However, with optimization $P_{i}$ changes in each iteration and one may achieve a faster decrease than in the non-optimized case. Fig~\ref{fig:CumulativeProbability} demonstrates the difference between these approaches: in the optimized case the probability of missing the ground state tends to zero faster than the non-optimized case. As we learn better $\kappa_{i}$ values after each iteration, the probability of finding the ground state is higher than in the non-optimized case.

\section{Conclusion}

We have studied the annealing of an archetypal transverse Ising spin (qubit) system coupled to an infinite heat bath with which it exchanges energy. We focused upon the probability of finding the ground state by the end of an experiment in {\emph{finite}} time, if repeated runs are permitted. While these repetitions may extend the overall run-time of the annealing procedure, the probability of missing the true ground state in all experiments diminishes exponentially. We have also analysed the role of the annealing time, $\tau$, and that of the site-dependent qubit-bath coupling strengths. However, it is left for future research to investigate the effects of graph theoretical quantities, such as adjacency structure, centrality, in-betweenness, etc, on the annealing procedure.

Focusing on a single experiment, we identified parameter ranges in which the environment can assist and improve the performance of the annealing. We have shown that the quench parameter, $\tau$, and the system-environment couplings, $\lbrace \kappa_{i} \rbrace$, can be optimized to improve the annealing process and keep the state of the quantum system close to the instantaneous ground state throughout the entire experiment. 

Finally, repeating the simulation multiple times, the system ground state can be identified quicker if the system-bath coupling strength is varied in a supervised way at each iteration, compared to if one repeats the simulation with the system-bath coupling strength maintained constant.

\bibliographystyle{apsrev4-1}

\begin{thebibliography}{58}%
\makeatletter
\providecommand \@ifxundefined [1]{%
 \@ifx{#1\undefined}
}%
\providecommand \@ifnum [1]{%
 \ifnum #1\expandafter \@firstoftwo
 \else \expandafter \@secondoftwo
 \fi
}%
\providecommand \@ifx [1]{%
 \ifx #1\expandafter \@firstoftwo
 \else \expandafter \@secondoftwo
 \fi
}%
\providecommand \natexlab [1]{#1}%
\providecommand \enquote  [1]{``#1''}%
\providecommand \bibnamefont  [1]{#1}%
\providecommand \bibfnamefont [1]{#1}%
\providecommand \citenamefont [1]{#1}%
\providecommand \href@noop [0]{\@secondoftwo}%
\providecommand \href [0]{\begingroup \@sanitize@url \@href}%
\providecommand \@href[1]{\@@startlink{#1}\@@href}%
\providecommand \@@href[1]{\endgroup#1\@@endlink}%
\providecommand \@sanitize@url [0]{\catcode `\\12\catcode `\$12\catcode
  `\&12\catcode `\#12\catcode `\^12\catcode `\_12\catcode `\%12\relax}%
\providecommand \@@startlink[1]{}%
\providecommand \@@endlink[0]{}%
\providecommand \url  [0]{\begingroup\@sanitize@url \@url }%
\providecommand \@url [1]{\endgroup\@href {#1}{\urlprefix }}%
\providecommand \urlprefix  [0]{URL }%
\providecommand \Eprint [0]{\href }%
\providecommand \doibase [0]{http://dx.doi.org/}%
\providecommand \selectlanguage [0]{\@gobble}%
\providecommand \bibinfo  [0]{\@secondoftwo}%
\providecommand \bibfield  [0]{\@secondoftwo}%
\providecommand \translation [1]{[#1]}%
\providecommand \BibitemOpen [0]{}%
\providecommand \bibitemStop [0]{}%
\providecommand \bibitemNoStop [0]{.\EOS\space}%
\providecommand \EOS [0]{\spacefactor3000\relax}%
\providecommand \BibitemShut  [1]{\csname bibitem#1\endcsname}%
\let\auto@bib@innerbib\@empty
\bibitem [{\citenamefont {Shor}(1994)}]{Shor1994}%
  \BibitemOpen
  \bibfield  {author} {\bibinfo {author} {\bibfnamefont {P.}~\bibnamefont
  {Shor}},\ }in\ \href {\doibase 10.1109/sfcs.1994.365700} {\emph {\bibinfo
  {booktitle} {Proceedings 35th Annual Symposium on Foundations of Computer
  Science}}}\ (\bibinfo  {publisher} {{IEEE} Comput. Soc. Press},\ \bibinfo
  {year} {1994})\BibitemShut {NoStop}%
\bibitem [{\citenamefont {Grover}(1997)}]{Grover1997}%
  \BibitemOpen
  \bibfield  {author} {\bibinfo {author} {\bibfnamefont {L.~K.}\ \bibnamefont
  {Grover}},\ }\href {\doibase 10.1103/physrevLetters79.325} {\bibfield
  {journal} {\bibinfo  {journal} {Physical Review Letters}\ }\textbf {\bibinfo
  {volume} {79}},\ \bibinfo {pages} {325} (\bibinfo {year} {1997})}\BibitemShut
  {NoStop}%
\bibitem [{\citenamefont {Lloyd}(1996)}]{Lloyd1996}%
  \BibitemOpen
  \bibfield  {author} {\bibinfo {author} {\bibfnamefont {S.}~\bibnamefont
  {Lloyd}},\ }\href {\doibase 10.1126/science.273.5278.1073} {\bibfield
  {journal} {\bibinfo  {journal} {Science}\ }\textbf {\bibinfo {volume}
  {273}},\ \bibinfo {pages} {1073} (\bibinfo {year} {1996})}\BibitemShut
  {NoStop}%
\bibitem [{\citenamefont {Abrams}\ and\ \citenamefont
  {Lloyd}(1999)}]{Abrams1999}%
  \BibitemOpen
  \bibfield  {author} {\bibinfo {author} {\bibfnamefont {D.~S.}\ \bibnamefont
  {Abrams}}\ and\ \bibinfo {author} {\bibfnamefont {S.}~\bibnamefont {Lloyd}},\
  }\href {\doibase 10.1103/PhysRevLett.83.5162} {\bibfield  {journal} {\bibinfo
   {journal} {Physical Review Letters}\ }\textbf {\bibinfo {volume} {83}},\
  \bibinfo {pages} {5162} (\bibinfo {year} {1999})}\BibitemShut {NoStop}%
\bibitem [{\citenamefont {Abrams}\ and\ \citenamefont
  {Lloyd}(1997)}]{Abrams1997}%
  \BibitemOpen
  \bibfield  {author} {\bibinfo {author} {\bibfnamefont {D.~S.}\ \bibnamefont
  {Abrams}}\ and\ \bibinfo {author} {\bibfnamefont {S.}~\bibnamefont {Lloyd}},\
  }\href {\doibase 10.1103/PhysRevLett.79.2586} {\bibfield  {journal} {\bibinfo
   {journal} {Physical Review Letters}\ }\textbf {\bibinfo {volume} {79}},\
  \bibinfo {pages} {2586} (\bibinfo {year} {1997})}\BibitemShut {NoStop}%
\bibitem [{\citenamefont {Cook}\ \emph {et~al.}(2011)\citenamefont {Cook},
  \citenamefont {Applegate}, \citenamefont {Bixby},\ and\ \citenamefont
  {Chv{\'{a}}tal}}]{Cook2011}%
  \BibitemOpen
  \bibfield  {author} {\bibinfo {author} {\bibfnamefont {W.~J.}\ \bibnamefont
  {Cook}}, \bibinfo {author} {\bibfnamefont {D.~L.}\ \bibnamefont {Applegate}},
  \bibinfo {author} {\bibfnamefont {R.~E.}\ \bibnamefont {Bixby}}, \ and\
  \bibinfo {author} {\bibfnamefont {V.}~\bibnamefont {Chv{\'{a}}tal}},\ }\href
  {\doibase 10.1515/9781400841103} {\emph {\bibinfo {title} {{T}he {T}raveling
  {S}alesman {P}roblem}}}\ (\bibinfo  {publisher} {Princeton University
  Press},\ \bibinfo {year} {2011})\BibitemShut {NoStop}%
\bibitem [{\citenamefont {Farhi}\ \emph {et~al.}(2000)\citenamefont {Farhi},
  \citenamefont {Goldstone}, \citenamefont {Gutmann},\ and\ \citenamefont
  {Sipser}}]{Farhi2000}%
  \BibitemOpen
  \bibfield  {author} {\bibinfo {author} {\bibfnamefont {E.}~\bibnamefont
  {Farhi}}, \bibinfo {author} {\bibfnamefont {J.}~\bibnamefont {Goldstone}},
  \bibinfo {author} {\bibfnamefont {S.}~\bibnamefont {Gutmann}}, \ and\
  \bibinfo {author} {\bibfnamefont {M.}~\bibnamefont {Sipser}},\ }\href@noop {}
  {\enquote {\bibinfo {title} {{Q}uantum {C}omputation by {A}diabatic
  {E}volution},}\ } (\bibinfo {year} {2000}),\ \Eprint
  {http://arxiv.org/abs/0001106} {arXiv:0001106 [quant-ph]} \BibitemShut
  {NoStop}%
\bibitem [{\citenamefont {Marto\ifmmode~\check{n}\else \v{n}\fi{}\'ak}\ \emph
  {et~al.}(2004)\citenamefont {Marto\ifmmode~\check{n}\else \v{n}\fi{}\'ak},
  \citenamefont {Santoro},\ and\ \citenamefont {Tosatti}}]{Martonak2004}%
  \BibitemOpen
  \bibfield  {author} {\bibinfo {author} {\bibfnamefont {R.}~\bibnamefont
  {Marto\ifmmode~\check{n}\else \v{n}\fi{}\'ak}}, \bibinfo {author}
  {\bibfnamefont {G.~E.}\ \bibnamefont {Santoro}}, \ and\ \bibinfo {author}
  {\bibfnamefont {E.}~\bibnamefont {Tosatti}},\ }\href {\doibase
  10.1103/PhysRevE.70.057701} {\bibfield  {journal} {\bibinfo  {journal}
  {Physical Review E}\ }\textbf {\bibinfo {volume} {70}},\ \bibinfo {pages}
  {057701} (\bibinfo {year} {2004})}\BibitemShut {NoStop}%
\bibitem [{\citenamefont {Pirchi}\ \emph {et~al.}(2011)\citenamefont {Pirchi},
  \citenamefont {Ziv}, \citenamefont {Riven}, \citenamefont {Cohen},
  \citenamefont {Zohar}, \citenamefont {Barak},\ and\ \citenamefont
  {Haran}}]{Pirchi2011}%
  \BibitemOpen
  \bibfield  {author} {\bibinfo {author} {\bibfnamefont {M.}~\bibnamefont
  {Pirchi}}, \bibinfo {author} {\bibfnamefont {G.}~\bibnamefont {Ziv}},
  \bibinfo {author} {\bibfnamefont {I.}~\bibnamefont {Riven}}, \bibinfo
  {author} {\bibfnamefont {S.~S.}\ \bibnamefont {Cohen}}, \bibinfo {author}
  {\bibfnamefont {N.}~\bibnamefont {Zohar}}, \bibinfo {author} {\bibfnamefont
  {Y.}~\bibnamefont {Barak}}, \ and\ \bibinfo {author} {\bibfnamefont
  {G.}~\bibnamefont {Haran}},\ }\href {\doibase 10.1038/ncomms1504} {\bibfield
  {journal} {\bibinfo  {journal} {Nature Communications}\ }\textbf {\bibinfo
  {volume} {2}} (\bibinfo {year} {2011}),\ 10.1038/ncomms1504}\BibitemShut
  {NoStop}%
\bibitem [{\citenamefont {Lucas}(2014)}]{Lucas2014}%
  \BibitemOpen
  \bibfield  {author} {\bibinfo {author} {\bibfnamefont {A.}~\bibnamefont
  {Lucas}},\ }\href {\doibase 10.3389/fphy.2014.00005} {\bibfield  {journal}
  {\bibinfo  {journal} {Frontiers in Physics}\ }\textbf {\bibinfo {volume}
  {2}},\ \bibinfo {pages} {5} (\bibinfo {year} {2014})}\BibitemShut {NoStop}%
\bibitem [{\citenamefont {Srinivasan}\ \emph {et~al.}(2018)\citenamefont
  {Srinivasan}, \citenamefont {Satyajit}, \citenamefont {Behera},\ and\
  \citenamefont {Panigrahi}}]{Srinivasan2018}%
  \BibitemOpen
  \bibfield  {author} {\bibinfo {author} {\bibfnamefont {K.}~\bibnamefont
  {Srinivasan}}, \bibinfo {author} {\bibfnamefont {S.}~\bibnamefont
  {Satyajit}}, \bibinfo {author} {\bibfnamefont {B.~K.}\ \bibnamefont
  {Behera}}, \ and\ \bibinfo {author} {\bibfnamefont {P.~K.}\ \bibnamefont
  {Panigrahi}},\ }\href@noop {} {\enquote {\bibinfo {title} {{E}fficient
  quantum algorithm for solving travelling salesman problem: {A}n {IBM} quantum
  experience},}\ } (\bibinfo {year} {2018}),\ \Eprint
  {http://arxiv.org/abs/1805.10928} {arXiv:1805.10928 [quant-ph]} \BibitemShut
  {NoStop}%
\bibitem [{\citenamefont {Warren}(2019)}]{Warren2019}%
  \BibitemOpen
  \bibfield  {author} {\bibinfo {author} {\bibfnamefont {R.~H.}\ \bibnamefont
  {Warren}},\ }\href {\doibase 10.1007/s42452-019-1829-x} {\bibfield  {journal}
  {\bibinfo  {journal} {{SN} Applied Sciences}\ }\textbf {\bibinfo {volume}
  {2}} (\bibinfo {year} {2019}),\ 10.1007/s42452-019-1829-x},\ \bibinfo {note}
  {article number: 75}\BibitemShut {NoStop}%
\bibitem [{\citenamefont {Dan}\ \emph {et~al.}(2020)\citenamefont {Dan},
  \citenamefont {Shimizu}, \citenamefont {Nishikawa}, \citenamefont {Bian},\
  and\ \citenamefont {Sato}}]{Dan2020}%
  \BibitemOpen
  \bibfield  {author} {\bibinfo {author} {\bibfnamefont {A.}~\bibnamefont
  {Dan}}, \bibinfo {author} {\bibfnamefont {R.}~\bibnamefont {Shimizu}},
  \bibinfo {author} {\bibfnamefont {T.}~\bibnamefont {Nishikawa}}, \bibinfo
  {author} {\bibfnamefont {S.}~\bibnamefont {Bian}}, \ and\ \bibinfo {author}
  {\bibfnamefont {T.}~\bibnamefont {Sato}},\ }in\ \href@noop {} {\emph
  {\bibinfo {booktitle} {Proceedings of the 57th ACM/EDAC/IEEE Design
  Automation Conference}}},\ \bibinfo {series and number} {DAC '20}\ (\bibinfo
  {publisher} {IEEE Press},\ \bibinfo {year} {2020})\BibitemShut {NoStop}%
\bibitem [{\citenamefont {Tan}\ \emph {et~al.}(2021)\citenamefont {Tan},
  \citenamefont {Lemonde}, \citenamefont {Thanasilp}, \citenamefont
  {Tangpanitanon},\ and\ \citenamefont {Angelakis}}]{Tan2021}%
  \BibitemOpen
  \bibfield  {author} {\bibinfo {author} {\bibfnamefont {B.}~\bibnamefont
  {Tan}}, \bibinfo {author} {\bibfnamefont {M.-A.}\ \bibnamefont {Lemonde}},
  \bibinfo {author} {\bibfnamefont {S.}~\bibnamefont {Thanasilp}}, \bibinfo
  {author} {\bibfnamefont {J.}~\bibnamefont {Tangpanitanon}}, \ and\ \bibinfo
  {author} {\bibfnamefont {D.~G.}\ \bibnamefont {Angelakis}},\ }\href {\doibase
  10.22331/q-2021-05-04-454} {\bibfield  {journal} {\bibinfo  {journal}
  {Quantum}\ }\textbf {\bibinfo {volume} {5}},\ \bibinfo {pages} {454}
  (\bibinfo {year} {2021})}\BibitemShut {NoStop}%
\bibitem [{\citenamefont {Barahona}(1982)}]{Barahona1982}%
  \BibitemOpen
  \bibfield  {author} {\bibinfo {author} {\bibfnamefont {F.}~\bibnamefont
  {Barahona}},\ }\href {\doibase 10.1088/0305-4470/15/10/028} {\bibfield
  {journal} {\bibinfo  {journal} {Journal of Physics A: Mathematical and
  General}\ }\textbf {\bibinfo {volume} {15}},\ \bibinfo {pages} {3241}
  (\bibinfo {year} {1982})}\BibitemShut {NoStop}%
\bibitem [{\citenamefont {Kato}(1950)}]{Kato1950}%
  \BibitemOpen
  \bibfield  {author} {\bibinfo {author} {\bibfnamefont {T.}~\bibnamefont
  {Kato}},\ }\href {\doibase 10.1143/JPSJ.5.435} {\bibfield  {journal}
  {\bibinfo  {journal} {Journal of the Physical Society of Japan}\ }\textbf
  {\bibinfo {volume} {5}},\ \bibinfo {pages} {435} (\bibinfo {year} {1950})},\
  \Eprint {http://arxiv.org/abs/https://doi.org/10.1143/JPSJ.5.435}
  {https://doi.org/10.1143/JPSJ.5.435} \BibitemShut {NoStop}%
\bibitem [{\citenamefont {Suzuki}(1970)}]{Suzuki1970}%
  \BibitemOpen
  \bibfield  {author} {\bibinfo {author} {\bibfnamefont {M.}~\bibnamefont
  {Suzuki}},\ }\href {\doibase 10.1143/PTP.43.882} {\bibfield  {journal}
  {\bibinfo  {journal} {Progress of Theoretical Physics}\ }\textbf {\bibinfo
  {volume} {43}},\ \bibinfo {pages} {882} (\bibinfo {year} {1970})},\ \Eprint
  {http://arxiv.org/abs/https://academic.oup.com/ptp/article-pdf/43/4/882/5467351/43-4-882.pdf}
  {https://academic.oup.com/ptp/article-pdf/43/4/882/5467351/43-4-882.pdf}
  \BibitemShut {NoStop}%
\bibitem [{\citenamefont {Salamon}\ and\ \citenamefont
  {Herman}(1978)}]{Salamon1978}%
  \BibitemOpen
  \bibfield  {author} {\bibinfo {author} {\bibfnamefont {M.~B.}\ \bibnamefont
  {Salamon}}\ and\ \bibinfo {author} {\bibfnamefont {R.~M.}\ \bibnamefont
  {Herman}},\ }\href {\doibase 10.1103/PhysRevLetters41.1506} {\bibfield
  {journal} {\bibinfo  {journal} {Physical Review Letters}\ }\textbf {\bibinfo
  {volume} {41}},\ \bibinfo {pages} {1506} (\bibinfo {year}
  {1978})}\BibitemShut {NoStop}%
\bibitem [{\citenamefont {Binder}\ \emph {et~al.}(2012)\citenamefont {Binder},
  \citenamefont {Binder}, \citenamefont {Ceperley}, \citenamefont {Hansen},
  \citenamefont {Kalos}, \citenamefont {Landau}, \citenamefont {Levesque},
  \citenamefont {M{\"u}ller-Krumbhaar}, \citenamefont {Stauffer},\ and\
  \citenamefont {Weis}}]{Binder2012}%
  \BibitemOpen
  \bibfield  {author} {\bibinfo {author} {\bibfnamefont {K.}~\bibnamefont
  {Binder}}, \bibinfo {author} {\bibfnamefont {K.}~\bibnamefont {Binder}},
  \bibinfo {author} {\bibfnamefont {D.}~\bibnamefont {Ceperley}}, \bibinfo
  {author} {\bibfnamefont {J.}~\bibnamefont {Hansen}}, \bibinfo {author}
  {\bibfnamefont {M.}~\bibnamefont {Kalos}}, \bibinfo {author} {\bibfnamefont
  {D.}~\bibnamefont {Landau}}, \bibinfo {author} {\bibfnamefont
  {D.}~\bibnamefont {Levesque}}, \bibinfo {author} {\bibfnamefont
  {H.}~\bibnamefont {M{\"u}ller-Krumbhaar}}, \bibinfo {author} {\bibfnamefont
  {D.}~\bibnamefont {Stauffer}}, \ and\ \bibinfo {author} {\bibfnamefont
  {J.}~\bibnamefont {Weis}},\ }\href
  {https://books.google.co.nz/books?id=nmfmCAAAQBAJ} {\emph {\bibinfo {title}
  {{M}onte {C}arlo {M}ethods in {S}tatistical {P}hysics}}},\ Topics in Current
  Physics\ (\bibinfo  {publisher} {Springer Berlin Heidelberg},\ \bibinfo
  {year} {2012})\BibitemShut {NoStop}%
\bibitem [{\citenamefont {Tredicce}\ \emph {et~al.}(2004)\citenamefont
  {Tredicce}, \citenamefont {Lippi}, \citenamefont {Mandel}, \citenamefont
  {Charasse}, \citenamefont {Chevalier},\ and\ \citenamefont
  {Picqu{\'e}}}]{Tredicce2004}%
  \BibitemOpen
  \bibfield  {author} {\bibinfo {author} {\bibfnamefont {J.~R.}\ \bibnamefont
  {Tredicce}}, \bibinfo {author} {\bibfnamefont {G.~L.}\ \bibnamefont {Lippi}},
  \bibinfo {author} {\bibfnamefont {P.}~\bibnamefont {Mandel}}, \bibinfo
  {author} {\bibfnamefont {B.}~\bibnamefont {Charasse}}, \bibinfo {author}
  {\bibfnamefont {A.}~\bibnamefont {Chevalier}}, \ and\ \bibinfo {author}
  {\bibfnamefont {B.}~\bibnamefont {Picqu{\'e}}},\ }\href {\doibase
  10.1119/1.1688783} {\bibfield  {journal} {\bibinfo  {journal} {American
  Journal of Physics}\ }\textbf {\bibinfo {volume} {72}},\ \bibinfo {pages}
  {799} (\bibinfo {year} {2004})},\ \Eprint
  {http://arxiv.org/abs/https://doi.org/10.1119/1.1688783}
  {https://doi.org/10.1119/1.1688783} \BibitemShut {NoStop}%
\bibitem [{\citenamefont {Biroli}\ \emph {et~al.}(2010)\citenamefont {Biroli},
  \citenamefont {Cugliandolo},\ and\ \citenamefont {Sicilia}}]{Biroli2010}%
  \BibitemOpen
  \bibfield  {author} {\bibinfo {author} {\bibfnamefont {G.}~\bibnamefont
  {Biroli}}, \bibinfo {author} {\bibfnamefont {L.~F.}\ \bibnamefont
  {Cugliandolo}}, \ and\ \bibinfo {author} {\bibfnamefont {A.}~\bibnamefont
  {Sicilia}},\ }\href {\doibase 10.1103/PhysRevE.81.050101} {\bibfield
  {journal} {\bibinfo  {journal} {Physical Review E}\ }\textbf {\bibinfo
  {volume} {81}},\ \bibinfo {pages} {050101(R)} (\bibinfo {year}
  {2010})}\BibitemShut {NoStop}%
\bibitem [{\citenamefont {Magalinskii}(1959)}]{Magalinskii1959}%
  \BibitemOpen
  \bibfield  {author} {\bibinfo {author} {\bibfnamefont {V.~B.}\ \bibnamefont
  {Magalinskii}},\ }\href {www.jetp.ras.ru/cgi-bin/e/index/r/36/6/p1942?a=list}
  {\bibfield  {journal} {\bibinfo  {journal} {JETP}\ }\textbf {\bibinfo
  {volume} {9}},\ \bibinfo {pages} {1382} (\bibinfo {year} {1959})}\BibitemShut
  {NoStop}%
\bibitem [{\citenamefont {Zwanzig}(1973)}]{Zwanzig1973}%
  \BibitemOpen
  \bibfield  {author} {\bibinfo {author} {\bibfnamefont {R.}~\bibnamefont
  {Zwanzig}},\ }\href {\doibase 10.1007/bf01008729} {\bibfield  {journal}
  {\bibinfo  {journal} {Journal of Statistical Physics}\ }\textbf {\bibinfo
  {volume} {9}},\ \bibinfo {pages} {215} (\bibinfo {year} {1973})}\BibitemShut
  {NoStop}%
\bibitem [{\citenamefont {Caldeira}\ and\ \citenamefont
  {Leggett}(1981)}]{Caldeira1981}%
  \BibitemOpen
  \bibfield  {author} {\bibinfo {author} {\bibfnamefont {A.~O.}\ \bibnamefont
  {Caldeira}}\ and\ \bibinfo {author} {\bibfnamefont {A.~J.}\ \bibnamefont
  {Leggett}},\ }\href {\doibase 10.1103/PhysRevLetters46.211} {\bibfield
  {journal} {\bibinfo  {journal} {Physical Review Letters}\ }\textbf {\bibinfo
  {volume} {46}},\ \bibinfo {pages} {211} (\bibinfo {year} {1981})}\BibitemShut
  {NoStop}%
\bibitem [{\citenamefont {Bu{\v{c}}a}\ \emph {et~al.}(2019)\citenamefont
  {Bu{\v{c}}a}, \citenamefont {Tindall},\ and\ \citenamefont
  {Jaksch}}]{Buca2019}%
  \BibitemOpen
  \bibfield  {author} {\bibinfo {author} {\bibfnamefont {B.}~\bibnamefont
  {Bu{\v{c}}a}}, \bibinfo {author} {\bibfnamefont {J.}~\bibnamefont {Tindall}},
  \ and\ \bibinfo {author} {\bibfnamefont {D.}~\bibnamefont {Jaksch}},\ }\href
  {\doibase 10.1038/s41467-019-09757-y} {\bibfield  {journal} {\bibinfo
  {journal} {Nature Communications}\ }\textbf {\bibinfo {volume} {10}}
  (\bibinfo {year} {2019}),\ 10.1038/s41467-019-09757-y}\BibitemShut {NoStop}%
\bibitem [{\citenamefont {Johnson}\ \emph {et~al.}(2011)\citenamefont
  {Johnson}, \citenamefont {Amin}, \citenamefont {Gildert}, \citenamefont
  {Lanting}, \citenamefont {Hamze}, \citenamefont {Dickson}, \citenamefont
  {Harris}, \citenamefont {Berkley}, \citenamefont {Johansson}, \citenamefont
  {Bunyk}, \citenamefont {Chapple}, \citenamefont {Enderud}, \citenamefont
  {Hilton}, \citenamefont {Karimi}, \citenamefont {Ladizinsky}, \citenamefont
  {Ladizinsky}, \citenamefont {Oh}, \citenamefont {Perminov}, \citenamefont
  {Rich}, \citenamefont {Thom}, \citenamefont {Tolkacheva}, \citenamefont
  {Truncik}, \citenamefont {Uchaikin}, \citenamefont {Wang}, \citenamefont
  {Wilson},\ and\ \citenamefont {Rose}}]{Johnson2011}%
  \BibitemOpen
  \bibfield  {author} {\bibinfo {author} {\bibfnamefont {M.~W.}\ \bibnamefont
  {Johnson}}, \bibinfo {author} {\bibfnamefont {M.~H.~S.}\ \bibnamefont
  {Amin}}, \bibinfo {author} {\bibfnamefont {S.}~\bibnamefont {Gildert}},
  \bibinfo {author} {\bibfnamefont {T.}~\bibnamefont {Lanting}}, \bibinfo
  {author} {\bibfnamefont {F.}~\bibnamefont {Hamze}}, \bibinfo {author}
  {\bibfnamefont {N.}~\bibnamefont {Dickson}}, \bibinfo {author} {\bibfnamefont
  {R.}~\bibnamefont {Harris}}, \bibinfo {author} {\bibfnamefont {A.~J.}\
  \bibnamefont {Berkley}}, \bibinfo {author} {\bibfnamefont {J.}~\bibnamefont
  {Johansson}}, \bibinfo {author} {\bibfnamefont {P.}~\bibnamefont {Bunyk}},
  \bibinfo {author} {\bibfnamefont {E.~M.}\ \bibnamefont {Chapple}}, \bibinfo
  {author} {\bibfnamefont {C.}~\bibnamefont {Enderud}}, \bibinfo {author}
  {\bibfnamefont {J.~P.}\ \bibnamefont {Hilton}}, \bibinfo {author}
  {\bibfnamefont {K.}~\bibnamefont {Karimi}}, \bibinfo {author} {\bibfnamefont
  {E.}~\bibnamefont {Ladizinsky}}, \bibinfo {author} {\bibfnamefont
  {N.}~\bibnamefont {Ladizinsky}}, \bibinfo {author} {\bibfnamefont
  {T.}~\bibnamefont {Oh}}, \bibinfo {author} {\bibfnamefont {I.}~\bibnamefont
  {Perminov}}, \bibinfo {author} {\bibfnamefont {C.}~\bibnamefont {Rich}},
  \bibinfo {author} {\bibfnamefont {M.~C.}\ \bibnamefont {Thom}}, \bibinfo
  {author} {\bibfnamefont {E.}~\bibnamefont {Tolkacheva}}, \bibinfo {author}
  {\bibfnamefont {C.~J.~S.}\ \bibnamefont {Truncik}}, \bibinfo {author}
  {\bibfnamefont {S.}~\bibnamefont {Uchaikin}}, \bibinfo {author}
  {\bibfnamefont {J.}~\bibnamefont {Wang}}, \bibinfo {author} {\bibfnamefont
  {B.}~\bibnamefont {Wilson}}, \ and\ \bibinfo {author} {\bibfnamefont
  {G.}~\bibnamefont {Rose}},\ }\href {\doibase 10.1038/nature10012} {\bibfield
  {journal} {\bibinfo  {journal} {Nature}\ }\textbf {\bibinfo {volume} {473}},\
  \bibinfo {pages} {194} (\bibinfo {year} {2011})}\BibitemShut {NoStop}%
\bibitem [{\citenamefont {Amin}\ \emph {et~al.}(2008)\citenamefont {Amin},
  \citenamefont {Love},\ and\ \citenamefont {Truncik}}]{Amin2008}%
  \BibitemOpen
  \bibfield  {author} {\bibinfo {author} {\bibfnamefont {M.~H.~S.}\
  \bibnamefont {Amin}}, \bibinfo {author} {\bibfnamefont {P.~J.}\ \bibnamefont
  {Love}}, \ and\ \bibinfo {author} {\bibfnamefont {C.~J.~S.}\ \bibnamefont
  {Truncik}},\ }\href {\doibase 10.1103/PhysRevLetters100.060503} {\bibfield
  {journal} {\bibinfo  {journal} {Physical Review Letters}\ }\textbf {\bibinfo
  {volume} {100}},\ \bibinfo {pages} {060503} (\bibinfo {year}
  {2008})}\BibitemShut {NoStop}%
\bibitem [{\citenamefont {Ashhab}(2014)}]{Ashhab2014}%
  \BibitemOpen
  \bibfield  {author} {\bibinfo {author} {\bibfnamefont {S.}~\bibnamefont
  {Ashhab}},\ }\href {\doibase 10.1103/PhysRevA.90.062120} {\bibfield
  {journal} {\bibinfo  {journal} {Physical Review A}\ }\textbf {\bibinfo
  {volume} {90}},\ \bibinfo {pages} {062120} (\bibinfo {year}
  {2014})}\BibitemShut {NoStop}%
\bibitem [{\citenamefont {Kechedzhi}\ and\ \citenamefont
  {Smelyanskiy}(2016)}]{Kechedzhi2016}%
  \BibitemOpen
  \bibfield  {author} {\bibinfo {author} {\bibfnamefont {K.}~\bibnamefont
  {Kechedzhi}}\ and\ \bibinfo {author} {\bibfnamefont {V.~N.}\ \bibnamefont
  {Smelyanskiy}},\ }\href {\doibase 10.1103/PhysRevX.6.021028} {\bibfield
  {journal} {\bibinfo  {journal} {Physical Review X}\ }\textbf {\bibinfo
  {volume} {6}},\ \bibinfo {pages} {021028} (\bibinfo {year}
  {2016})}\BibitemShut {NoStop}%
\bibitem [{\citenamefont {Keck}\ \emph {et~al.}(2017)\citenamefont {Keck},
  \citenamefont {Montangero}, \citenamefont {Santoro}, \citenamefont {Fazio},\
  and\ \citenamefont {Rossini}}]{Keck2017}%
  \BibitemOpen
  \bibfield  {author} {\bibinfo {author} {\bibfnamefont {M.}~\bibnamefont
  {Keck}}, \bibinfo {author} {\bibfnamefont {S.}~\bibnamefont {Montangero}},
  \bibinfo {author} {\bibfnamefont {G.~E.}\ \bibnamefont {Santoro}}, \bibinfo
  {author} {\bibfnamefont {R.}~\bibnamefont {Fazio}}, \ and\ \bibinfo {author}
  {\bibfnamefont {D.}~\bibnamefont {Rossini}},\ }\href {\doibase
  10.1088/1367-2630/aa8cef} {\bibfield  {journal} {\bibinfo  {journal} {New
  Journal of Physics}\ }\textbf {\bibinfo {volume} {19}},\ \bibinfo {pages}
  {113029} (\bibinfo {year} {2017})}\BibitemShut {NoStop}%
\bibitem [{\citenamefont {Arceci}\ \emph {et~al.}(2017)\citenamefont {Arceci},
  \citenamefont {Barbarino}, \citenamefont {Fazio},\ and\ \citenamefont
  {Santoro}}]{Arceci2017}%
  \BibitemOpen
  \bibfield  {author} {\bibinfo {author} {\bibfnamefont {L.}~\bibnamefont
  {Arceci}}, \bibinfo {author} {\bibfnamefont {S.}~\bibnamefont {Barbarino}},
  \bibinfo {author} {\bibfnamefont {R.}~\bibnamefont {Fazio}}, \ and\ \bibinfo
  {author} {\bibfnamefont {G.~E.}\ \bibnamefont {Santoro}},\ }\href {\doibase
  10.1103/PhysRevB.96.054301} {\bibfield  {journal} {\bibinfo  {journal}
  {Physical Review B}\ }\textbf {\bibinfo {volume} {96}},\ \bibinfo {pages}
  {054301} (\bibinfo {year} {2017})}\BibitemShut {NoStop}%
\bibitem [{\citenamefont {Arceci}\ \emph {et~al.}(2018)\citenamefont {Arceci},
  \citenamefont {Barbarino}, \citenamefont {Rossini},\ and\ \citenamefont
  {Santoro}}]{Arceci2018}%
  \BibitemOpen
  \bibfield  {author} {\bibinfo {author} {\bibfnamefont {L.}~\bibnamefont
  {Arceci}}, \bibinfo {author} {\bibfnamefont {S.}~\bibnamefont {Barbarino}},
  \bibinfo {author} {\bibfnamefont {D.}~\bibnamefont {Rossini}}, \ and\
  \bibinfo {author} {\bibfnamefont {G.~E.}\ \bibnamefont {Santoro}},\ }\href
  {\doibase 10.1103/PhysRevB.98.064307} {\bibfield  {journal} {\bibinfo
  {journal} {Physical Review B}\ }\textbf {\bibinfo {volume} {98}},\ \bibinfo
  {pages} {064307} (\bibinfo {year} {2018})}\BibitemShut {NoStop}%
\bibitem [{\citenamefont {Passarelli}\ \emph {et~al.}(2018)\citenamefont
  {Passarelli}, \citenamefont {De~Filippis}, \citenamefont {Cataudella},\ and\
  \citenamefont {Lucignano}}]{Passarelli2018}%
  \BibitemOpen
  \bibfield  {author} {\bibinfo {author} {\bibfnamefont {G.}~\bibnamefont
  {Passarelli}}, \bibinfo {author} {\bibfnamefont {G.}~\bibnamefont
  {De~Filippis}}, \bibinfo {author} {\bibfnamefont {V.}~\bibnamefont
  {Cataudella}}, \ and\ \bibinfo {author} {\bibfnamefont {P.}~\bibnamefont
  {Lucignano}},\ }\href {\doibase 10.1103/PhysRevA.97.022319} {\bibfield
  {journal} {\bibinfo  {journal} {Physical Review A}\ }\textbf {\bibinfo
  {volume} {97}},\ \bibinfo {pages} {022319} (\bibinfo {year}
  {2018})}\BibitemShut {NoStop}%
\bibitem [{\citenamefont {Lee}\ \emph {et~al.}(2019)\citenamefont {Lee},
  \citenamefont {Najafabadi}, \citenamefont {Schumayer}, \citenamefont {Kwek},\
  and\ \citenamefont {Hutchinson}}]{Lee2019}%
  \BibitemOpen
  \bibfield  {author} {\bibinfo {author} {\bibfnamefont {C.~K.}\ \bibnamefont
  {Lee}}, \bibinfo {author} {\bibfnamefont {M.~S.}\ \bibnamefont {Najafabadi}},
  \bibinfo {author} {\bibfnamefont {D.}~\bibnamefont {Schumayer}}, \bibinfo
  {author} {\bibfnamefont {L.~C.}\ \bibnamefont {Kwek}}, \ and\ \bibinfo
  {author} {\bibfnamefont {D.~A.~W.}\ \bibnamefont {Hutchinson}},\ }\href
  {\doibase 10.1038/s41598-019-45496-2} {\bibfield  {journal} {\bibinfo
  {journal} {Scientific Reports}\ }\textbf {\bibinfo {volume} {9}} (\bibinfo
  {year} {2019}),\ 10.1038/s41598-019-45496-2}\BibitemShut {NoStop}%
\bibitem [{\citenamefont {Oh}\ \emph {et~al.}(2019)\citenamefont {Oh},
  \citenamefont {Coker},\ and\ \citenamefont {Hutchinson}}]{Oh2019}%
  \BibitemOpen
  \bibfield  {author} {\bibinfo {author} {\bibfnamefont {S.~A.}\ \bibnamefont
  {Oh}}, \bibinfo {author} {\bibfnamefont {D.~F.}\ \bibnamefont {Coker}}, \
  and\ \bibinfo {author} {\bibfnamefont {D.~A.~W.}\ \bibnamefont
  {Hutchinson}},\ }\href {\doibase 10.1063/1.5048058} {\bibfield  {journal}
  {\bibinfo  {journal} {The Journal of Chemical Physics}\ }\textbf {\bibinfo
  {volume} {150}},\ \bibinfo {pages} {085102} (\bibinfo {year} {2019})},\
  \Eprint {http://arxiv.org/abs/https://doi.org/10.1063/1.5048058}
  {https://doi.org/10.1063/1.5048058} \BibitemShut {NoStop}%
\bibitem [{\citenamefont {Oh}\ \emph {et~al.}(2020)\citenamefont {Oh},
  \citenamefont {Coker},\ and\ \citenamefont {Hutchinson}}]{Oh2020}%
  \BibitemOpen
  \bibfield  {author} {\bibinfo {author} {\bibfnamefont {S.~A.}\ \bibnamefont
  {Oh}}, \bibinfo {author} {\bibfnamefont {D.~F.}\ \bibnamefont {Coker}}, \
  and\ \bibinfo {author} {\bibfnamefont {D.~A.}\ \bibnamefont {Hutchinson}},\
  }\href {\doibase 10.1039/C9FD00081J} {\bibfield  {journal} {\bibinfo
  {journal} {Faraday Discuss.}\ }\textbf {\bibinfo {volume} {221}},\ \bibinfo
  {pages} {59} (\bibinfo {year} {2020})}\BibitemShut {NoStop}%
\bibitem [{\citenamefont {Gaspard}\ and\ \citenamefont
  {Nagaoka}(1999)}]{Gaspard1999}%
  \BibitemOpen
  \bibfield  {author} {\bibinfo {author} {\bibfnamefont {P.}~\bibnamefont
  {Gaspard}}\ and\ \bibinfo {author} {\bibfnamefont {M.}~\bibnamefont
  {Nagaoka}},\ }\href {\doibase 10.1063/1.479867} {\bibfield  {journal}
  {\bibinfo  {journal} {The Journal of Chemical Physics}\ }\textbf {\bibinfo
  {volume} {111}},\ \bibinfo {pages} {5668} (\bibinfo {year} {1999})},\ \Eprint
  {http://arxiv.org/abs/https://doi.org/10.1063/1.479867}
  {https://doi.org/10.1063/1.479867} \BibitemShut {NoStop}%
\bibitem [{\citenamefont {Nalbach}\ and\ \citenamefont
  {Thorwart}(2009)}]{Nalbach2009}%
  \BibitemOpen
  \bibfield  {author} {\bibinfo {author} {\bibfnamefont {P.}~\bibnamefont
  {Nalbach}}\ and\ \bibinfo {author} {\bibfnamefont {M.}~\bibnamefont
  {Thorwart}},\ }\href {\doibase 10.1103/PhysRevLetters103.220401} {\bibfield
  {journal} {\bibinfo  {journal} {Physical Review Letters}\ }\textbf {\bibinfo
  {volume} {103}},\ \bibinfo {pages} {220401} (\bibinfo {year}
  {2009})}\BibitemShut {NoStop}%
\bibitem [{\citenamefont {May}(2011)}]{May2011}%
  \BibitemOpen
  \bibfield  {author} {\bibinfo {author} {\bibfnamefont {V.}~\bibnamefont
  {May}},\ }\href@noop {} {\emph {\bibinfo {title} {Charge and energy transfer
  dynamics in molecular systems}}}\ (\bibinfo  {publisher} {Wiley-VCH},\
  \bibinfo {address} {Weinheim},\ \bibinfo {year} {2011})\BibitemShut {NoStop}%
\bibitem [{\citenamefont {Nalbach}(2014)}]{Nalbach2014}%
  \BibitemOpen
  \bibfield  {author} {\bibinfo {author} {\bibfnamefont {P.}~\bibnamefont
  {Nalbach}},\ }\href {\doibase 10.1103/PhysRevA.90.042112} {\bibfield
  {journal} {\bibinfo  {journal} {Physical Review A}\ }\textbf {\bibinfo
  {volume} {90}},\ \bibinfo {pages} {042112} (\bibinfo {year}
  {2014})}\BibitemShut {NoStop}%
\bibitem [{\citenamefont {Yoshimura}\ and\ \citenamefont
  {Freericks}(2015)}]{Yoshimura2015}%
  \BibitemOpen
  \bibfield  {author} {\bibinfo {author} {\bibfnamefont {B.}~\bibnamefont
  {Yoshimura}}\ and\ \bibinfo {author} {\bibfnamefont {J.~K.}\ \bibnamefont
  {Freericks}},\ }\href {\doibase 10.3389/fphy.2015.00085} {\bibfield
  {journal} {\bibinfo  {journal} {Frontiers in Physics}\ }\textbf {\bibinfo
  {volume} {3}},\ \bibinfo {pages} {85} (\bibinfo {year} {2015})}\BibitemShut
  {NoStop}%
\bibitem [{\citenamefont {Javanbakht}\ \emph {et~al.}(2015)\citenamefont
  {Javanbakht}, \citenamefont {Nalbach},\ and\ \citenamefont
  {Thorwart}}]{Javanbakht2015}%
  \BibitemOpen
  \bibfield  {author} {\bibinfo {author} {\bibfnamefont {S.}~\bibnamefont
  {Javanbakht}}, \bibinfo {author} {\bibfnamefont {P.}~\bibnamefont {Nalbach}},
  \ and\ \bibinfo {author} {\bibfnamefont {M.}~\bibnamefont {Thorwart}},\
  }\href {\doibase 10.1103/PhysRevA.91.052103} {\bibfield  {journal} {\bibinfo
  {journal} {Physical Review A}\ }\textbf {\bibinfo {volume} {91}},\ \bibinfo
  {pages} {052103} (\bibinfo {year} {2015})}\BibitemShut {NoStop}%
\bibitem [{\citenamefont {Watabe}\ \emph {et~al.}(2020)\citenamefont {Watabe},
  \citenamefont {Seki},\ and\ \citenamefont {Kawabata}}]{Watabe2020}%
  \BibitemOpen
  \bibfield  {author} {\bibinfo {author} {\bibfnamefont {S.}~\bibnamefont
  {Watabe}}, \bibinfo {author} {\bibfnamefont {Y.}~\bibnamefont {Seki}}, \ and\
  \bibinfo {author} {\bibfnamefont {S.}~\bibnamefont {Kawabata}},\ }\href
  {\doibase 10.1038/s41598-019-56758-4} {\bibfield  {journal} {\bibinfo
  {journal} {Scientific Reports}\ }\textbf {\bibinfo {volume} {10}} (\bibinfo
  {year} {2020}),\ 10.1038/s41598-019-56758-4}\BibitemShut {NoStop}%
\bibitem [{\citenamefont {Bender}\ and\ \citenamefont
  {Canfield}(1978)}]{Bender1978}%
  \BibitemOpen
  \bibfield  {author} {\bibinfo {author} {\bibfnamefont {E.~A.}\ \bibnamefont
  {Bender}}\ and\ \bibinfo {author} {\bibfnamefont {E.}~\bibnamefont
  {Canfield}},\ }\href {\doibase https://doi.org/10.1016/0097-3165(78)90059-6}
  {\bibfield  {journal} {\bibinfo  {journal} {Journal of Combinatorial Theory,
  Series A}\ }\textbf {\bibinfo {volume} {24}},\ \bibinfo {pages} {296}
  (\bibinfo {year} {1978})}\BibitemShut {NoStop}%
\bibitem [{\citenamefont {Cohen-Tannoudji}\ \emph {et~al.}(1992)\citenamefont
  {Cohen-Tannoudji}, \citenamefont {Grynberg},\ and\ \citenamefont
  {Dupont-Roc}}]{Tannoudji1992}%
  \BibitemOpen
  \bibfield  {author} {\bibinfo {author} {\bibfnamefont {C.}~\bibnamefont
  {Cohen-Tannoudji}}, \bibinfo {author} {\bibfnamefont {G.}~\bibnamefont
  {Grynberg}}, \ and\ \bibinfo {author} {\bibfnamefont {J.}~\bibnamefont
  {Dupont-Roc}},\ }\href@noop {} {\emph {\bibinfo {title} {{A}tom-{P}hoton
  {I}nteractions: {B}asic {P}rocesses and {A}pplications}}}\ (\bibinfo
  {publisher} {Wiley},\ \bibinfo {address} {New York},\ \bibinfo {year}
  {1992})\BibitemShut {NoStop}%
\bibitem [{\citenamefont {Breuer}\ \emph {et~al.}(2002)\citenamefont {Breuer},
  \citenamefont {Petruccione},\ and\ \citenamefont {Petruccione}}]{Breuer2002}%
  \BibitemOpen
  \bibfield  {author} {\bibinfo {author} {\bibfnamefont {H.}~\bibnamefont
  {Breuer}}, \bibinfo {author} {\bibfnamefont {F.}~\bibnamefont {Petruccione}},
  \ and\ \bibinfo {author} {\bibfnamefont {S.}~\bibnamefont {Petruccione}},\
  }\href@noop {} {\emph {\bibinfo {title} {{T}he {T}heory of {O}pen {Q}uantum
  {S}ystems}}}\ (\bibinfo  {publisher} {Oxford University Press},\ \bibinfo
  {year} {2002})\BibitemShut {NoStop}%
\bibitem [{\citenamefont {Johansson}\ \emph {et~al.}(2013)\citenamefont
  {Johansson}, \citenamefont {Nation},\ and\ \citenamefont
  {Nori}}]{Johansson2013}%
  \BibitemOpen
  \bibfield  {author} {\bibinfo {author} {\bibfnamefont {J.}~\bibnamefont
  {Johansson}}, \bibinfo {author} {\bibfnamefont {P.}~\bibnamefont {Nation}}, \
  and\ \bibinfo {author} {\bibfnamefont {F.}~\bibnamefont {Nori}},\ }\href
  {\doibase https://doi.org/10.1016/j.cpc.2012.11.019} {\bibfield  {journal}
  {\bibinfo  {journal} {Computer Physics Communications}\ }\textbf {\bibinfo
  {volume} {184}},\ \bibinfo {pages} {1234} (\bibinfo {year}
  {2013})}\BibitemShut {NoStop}%
\bibitem [{\citenamefont {Lidar}(2020)}]{Lidar2020}%
  \BibitemOpen
  \bibfield  {author} {\bibinfo {author} {\bibfnamefont {D.~A.}\ \bibnamefont
  {Lidar}},\ }\href@noop {} {\enquote {\bibinfo {title} {{L}ecture {N}otes on
  the {T}heory of {O}pen {Q}uantum {S}ystems},}\ } (\bibinfo {year} {2020}),\
  \Eprint {http://arxiv.org/abs/1902.00967} {arXiv:1902.00967 [quant-ph]}
  \BibitemShut {NoStop}%
\bibitem [{\citenamefont {Yan}\ \emph {et~al.}(2016)\citenamefont {Yan},
  \citenamefont {Gustavsson}, \citenamefont {Kamal}, \citenamefont {Birenbaum},
  \citenamefont {Sears}, \citenamefont {Hover}, \citenamefont {Gudmundsen},
  \citenamefont {Rosenberg}, \citenamefont {Samach}, \citenamefont {Weber},
  \citenamefont {Yoder}, \citenamefont {Orlando}, \citenamefont {Clarke},
  \citenamefont {Kerman},\ and\ \citenamefont {Oliver}}]{Yan2016}%
  \BibitemOpen
  \bibfield  {author} {\bibinfo {author} {\bibfnamefont {F.}~\bibnamefont
  {Yan}}, \bibinfo {author} {\bibfnamefont {S.}~\bibnamefont {Gustavsson}},
  \bibinfo {author} {\bibfnamefont {A.}~\bibnamefont {Kamal}}, \bibinfo
  {author} {\bibfnamefont {J.}~\bibnamefont {Birenbaum}}, \bibinfo {author}
  {\bibfnamefont {A.~P.}\ \bibnamefont {Sears}}, \bibinfo {author}
  {\bibfnamefont {D.}~\bibnamefont {Hover}}, \bibinfo {author} {\bibfnamefont
  {T.~J.}\ \bibnamefont {Gudmundsen}}, \bibinfo {author} {\bibfnamefont
  {D.}~\bibnamefont {Rosenberg}}, \bibinfo {author} {\bibfnamefont
  {G.}~\bibnamefont {Samach}}, \bibinfo {author} {\bibfnamefont
  {S.}~\bibnamefont {Weber}}, \bibinfo {author} {\bibfnamefont {J.~L.}\
  \bibnamefont {Yoder}}, \bibinfo {author} {\bibfnamefont {T.~P.}\ \bibnamefont
  {Orlando}}, \bibinfo {author} {\bibfnamefont {J.}~\bibnamefont {Clarke}},
  \bibinfo {author} {\bibfnamefont {A.~J.}\ \bibnamefont {Kerman}}, \ and\
  \bibinfo {author} {\bibfnamefont {W.~D.}\ \bibnamefont {Oliver}},\ }\href
  {\doibase 10.1038/ncomms12964} {\bibfield  {journal} {\bibinfo  {journal}
  {Nature Communications}\ }\textbf {\bibinfo {volume} {7}} (\bibinfo {year}
  {2016}),\ 10.1038/ncomms12964}\BibitemShut {NoStop}%
\bibitem [{\citenamefont {Astafiev}\ \emph {et~al.}(2004)\citenamefont
  {Astafiev}, \citenamefont {Pashkin}, \citenamefont {Nakamura}, \citenamefont
  {Yamamoto},\ and\ \citenamefont {Tsai}}]{Astafiev2004}%
  \BibitemOpen
  \bibfield  {author} {\bibinfo {author} {\bibfnamefont {O.}~\bibnamefont
  {Astafiev}}, \bibinfo {author} {\bibfnamefont {Y.~A.}\ \bibnamefont
  {Pashkin}}, \bibinfo {author} {\bibfnamefont {Y.}~\bibnamefont {Nakamura}},
  \bibinfo {author} {\bibfnamefont {T.}~\bibnamefont {Yamamoto}}, \ and\
  \bibinfo {author} {\bibfnamefont {J.~S.}\ \bibnamefont {Tsai}},\ }\href
  {\doibase 10.1103/PhysRevLett.93.267007} {\bibfield  {journal} {\bibinfo
  {journal} {Phys. Rev. Lett.}\ }\textbf {\bibinfo {volume} {93}},\ \bibinfo
  {pages} {267007} (\bibinfo {year} {2004})}\BibitemShut {NoStop}%
\bibitem [{\citenamefont {Shnirman}\ \emph {et~al.}(2005)\citenamefont
  {Shnirman}, \citenamefont {Sch\"on}, \citenamefont {Martin},\ and\
  \citenamefont {Makhlin}}]{Shnirman2005}%
  \BibitemOpen
  \bibfield  {author} {\bibinfo {author} {\bibfnamefont {A.}~\bibnamefont
  {Shnirman}}, \bibinfo {author} {\bibfnamefont {G.}~\bibnamefont {Sch\"on}},
  \bibinfo {author} {\bibfnamefont {I.}~\bibnamefont {Martin}}, \ and\ \bibinfo
  {author} {\bibfnamefont {Y.}~\bibnamefont {Makhlin}},\ }\href {\doibase
  10.1103/PhysRevLett.94.127002} {\bibfield  {journal} {\bibinfo  {journal}
  {Phys. Rev. Lett.}\ }\textbf {\bibinfo {volume} {94}},\ \bibinfo {pages}
  {127002} (\bibinfo {year} {2005})}\BibitemShut {NoStop}%
\bibitem [{\citenamefont {Johansson}\ \emph {et~al.}(2012)\citenamefont
  {Johansson}, \citenamefont {Nation},\ and\ \citenamefont
  {Nori}}]{Johansson2012}%
  \BibitemOpen
  \bibfield  {author} {\bibinfo {author} {\bibfnamefont {J.}~\bibnamefont
  {Johansson}}, \bibinfo {author} {\bibfnamefont {P.}~\bibnamefont {Nation}}, \
  and\ \bibinfo {author} {\bibfnamefont {F.}~\bibnamefont {Nori}},\ }\href
  {\doibase https://doi.org/10.1016/j.cpc.2012.02.021} {\bibfield  {journal}
  {\bibinfo  {journal} {Computer Physics Communications}\ }\textbf {\bibinfo
  {volume} {183}},\ \bibinfo {pages} {1760} (\bibinfo {year}
  {2012})}\BibitemShut {NoStop}%
\bibitem [{\citenamefont {Albash}\ \emph {et~al.}(2012)\citenamefont {Albash},
  \citenamefont {Boixo}, \citenamefont {Lidar},\ and\ \citenamefont
  {Zanardi}}]{Albash2012}%
  \BibitemOpen
  \bibfield  {author} {\bibinfo {author} {\bibfnamefont {T.}~\bibnamefont
  {Albash}}, \bibinfo {author} {\bibfnamefont {S.}~\bibnamefont {Boixo}},
  \bibinfo {author} {\bibfnamefont {D.~A.}\ \bibnamefont {Lidar}}, \ and\
  \bibinfo {author} {\bibfnamefont {P.}~\bibnamefont {Zanardi}},\ }\href
  {\doibase 10.1088/1367-2630/14/12/123016} {\bibfield  {journal} {\bibinfo
  {journal} {New Journal of Physics}\ }\textbf {\bibinfo {volume} {14}},\
  \bibinfo {pages} {123016} (\bibinfo {year} {2012})}\BibitemShut {NoStop}%
\bibitem [{\citenamefont {Cattaneo}\ \emph {et~al.}(2019)\citenamefont
  {Cattaneo}, \citenamefont {Giorgi}, \citenamefont {Maniscalco},\ and\
  \citenamefont {Zambrini}}]{Cattaneo2019}%
  \BibitemOpen
  \bibfield  {author} {\bibinfo {author} {\bibfnamefont {M.}~\bibnamefont
  {Cattaneo}}, \bibinfo {author} {\bibfnamefont {G.~L.}\ \bibnamefont
  {Giorgi}}, \bibinfo {author} {\bibfnamefont {S.}~\bibnamefont {Maniscalco}},
  \ and\ \bibinfo {author} {\bibfnamefont {R.}~\bibnamefont {Zambrini}},\
  }\href {\doibase 10.1088/1367-2630/ab54ac} {\bibfield  {journal} {\bibinfo
  {journal} {New Journal of Physics}\ }\textbf {\bibinfo {volume} {21}},\
  \bibinfo {pages} {113045} (\bibinfo {year} {2019})}\BibitemShut {NoStop}%
\bibitem [{\citenamefont {Jozsa}(1994)}]{Jozsa1994}%
  \BibitemOpen
  \bibfield  {author} {\bibinfo {author} {\bibfnamefont {R.}~\bibnamefont
  {Jozsa}},\ }\href {\doibase 10.1080/09500349414552171} {\bibfield  {journal}
  {\bibinfo  {journal} {Journal of Modern Optics}\ }\textbf {\bibinfo {volume}
  {41}},\ \bibinfo {pages} {2315} (\bibinfo {year} {1994})}\BibitemShut
  {NoStop}%
\bibitem [{\citenamefont {Miszczak}\ \emph {et~al.}(2008)\citenamefont
  {Miszczak}, \citenamefont {Pucha{\l}a}, \citenamefont {Horodecki},
  \citenamefont {Uhlmann},\ and\ \citenamefont {{\.
  Z}yczkowski}}]{Miszczak2008}%
  \BibitemOpen
  \bibfield  {author} {\bibinfo {author} {\bibfnamefont {J.~A.}\ \bibnamefont
  {Miszczak}}, \bibinfo {author} {\bibfnamefont {Z.}~\bibnamefont
  {Pucha{\l}a}}, \bibinfo {author} {\bibfnamefont {P.}~\bibnamefont
  {Horodecki}}, \bibinfo {author} {\bibfnamefont {A.}~\bibnamefont {Uhlmann}},
  \ and\ \bibinfo {author} {\bibfnamefont {K.}~\bibnamefont {{\.
  Z}yczkowski}},\ }\href@noop {} {\enquote {\bibinfo {title} {Sub-- and
  super-fidelity as bounds for quantum fidelity},}\ } (\bibinfo {year}
  {2008}),\ \Eprint {http://arxiv.org/abs/0805.2037} {arXiv:0805.2037
  [quant-ph]} \BibitemShut {NoStop}%
\bibitem [{Note1()}]{Note1}%
  \BibitemOpen
  \bibinfo {note} {Each simulation consists of two parts: one solves the
  Bloch-Redfield master equation and measure the fidelity between two given
  states, and the other optimizes the target function $y = 1 - \protect
  \qopname \relax m{max}{\protect \tmspace -\thinmuskip {.1667em}\protect \bigl
  ( \protect \text {fidelity} \protect \bigr )}^{2}$ with the constraint $0 <
  \kappa $ . We rely on the {\protect \texttt {python3}} optimization module
  {\protect \texttt {scipy.optimize}} implementing the Nelder-Mead algorithm
  \cite {Nelder1965}. A run starts with an initial guess for $\kappa $,
  measures $F$ as in Eq.~\protect \textup {\hbox {\mathsurround \z@ \protect
  \normalfont (\ignorespaces \ref {eq:MiszczakFidelity}\unskip \@@italiccorr
  )}}, and then $y$ is evaluated.}\BibitemShut {Stop}%
\bibitem [{\citenamefont {Nelder}\ and\ \citenamefont
  {Mead}(1965)}]{Nelder1965}%
  \BibitemOpen
  \bibfield  {author} {\bibinfo {author} {\bibfnamefont {J.~A.}\ \bibnamefont
  {Nelder}}\ and\ \bibinfo {author} {\bibfnamefont {R.}~\bibnamefont {Mead}},\
  }\href {\doibase 10.1093/comjnl/7.4.308} {\bibfield  {journal} {\bibinfo
  {journal} {The Computer Journal}\ }\textbf {\bibinfo {volume} {7}},\ \bibinfo
  {pages} {308} (\bibinfo {year} {1965})}\BibitemShut {NoStop}%
\end{thebibliography}

%

\end{document}